\documentclass{aa}  

\usepackage{graphicx}
\usepackage{txfonts}
\newcommand{\Bdip}{\ensuremath{B_{\rm \,dip}}}

\newcommand{\Bupp}{\ensuremath{B_{\rm \,dip}^{\rm (max)}}}
\newcommand{\Nupp}{\ensuremath{N_{\rm \,dip}^{\rm (max)}}}

\newcommand{\bz}{\ensuremath{\langle B_z\rangle}}
\newcommand{\nz}{\ensuremath{\langle N_z\rangle}}
\newcommand{\snr}{{\it S/N}}

\bibpunct{(}{)}{;}{a}{}{,}

\newcommand{\pv}{\ensuremath{P_V}}
\newcommand{\nnv}{\ensuremath{N_V}}
\newcommand{\geff}{\ensuremath{g_{\rm eff}}}
\newcommand{\cz}{\ensuremath{c_z}}
\newcommand{\HeII}{He\,{\sc II}\,$\lambda4686$}

\def\gtrsim{\mathrel{\hbox{\rlap{\hbox{\lower4pt\hbox{$\sim$}}}\hbox{$>$}}}}
\def\ltsim{\mathrel{\hbox{\rlap{\hbox{\lower4pt\hbox{$\sim$}}}\hbox{$<$}}}}
\begin{document}

   \title{A search for strong magnetic fields in\\ massive and very massive stars in the Magellanic Clouds}

   \subtitle{}

\author{
       S.~Bagnulo         \inst{1}
       \and
       G.A.~Wade          \inst{2}
       \and
       Y.~Naz\'e          \inst{3} \thanks{FNRS senior research associate} 
       \and
       J.H. Grunhut       \inst{4}
       \and
       M.~E.~Shultz       \inst{5}
       \and
       D.J. Asher         \inst{1}
       \and
       P.A. Crowther      \inst{6}
       \and
       C.J. Evans         \inst{7}
       \and
       A.~David-Uraz      \inst{5}
       \and
       I.D.~Howarth       \inst{8}
       \and
       N.~Morrell         \inst{9}
       \and
       M.S.~Munoz        \inst{10}
       \and
       C.~Neiner          \inst{11}
       \and
       J.~Puls           \inst{12}
       \and 
       M.K.~Szyma\'nski   \inst{13}
       \and
       J.S.~Vink         \inst{1}
}
\institute{
           Armagh Observatory and Planetarium, College Hill, Armagh BT61 9DG, UK   \email{stefano.bagnulo@armagh.ac.uk}
           \and
           Department of Physics and Space Science, Royal Military College of Canada, P.O. Box 17000,
           Station Forces, ON, Canada K7K 4B4  
           \and
           Institut d'Astrophysique et de G\'eophysique, Universit\'e de Li\`ege,
           Quartier Agora (B5c), All\'ee du 6 Ao\^ut 19c, B-4000 Sart Tilman, Li\`ege, Belgium  
           \and
           European Southern Observatory, Karl-Schwarzschild-Str. 2, D-85748 Garching, Germany  
           \and
           Department of Physics and Astronomy, University of Delaware, Newark, DE 19716, USA
           \and
           Department of Physics and Astronomy, University of Sheffield, Hicks Building, Hounsfield Road, Sheffield S3 7RH, UK 
           \and
           UK Astronomy Technology Centre, Royal Observatory, Blackford Hill, Edinburgh EH9 3HJ, UK
           \and
           Department of Physics and Astronomy, University College London, Gower Street, London WC1E 6BT, UK
           \and
           Las Campanas Observatory, Carnegie Observatories, Casilla 601, La Serena, Chile
           \and
           Department of Physics, Engineering Physics and Astronomy, Queen's University,
           64 Bader lane, Kingston, K7L 3N6, ON Canada
           \and
           LESIA, Paris Observatory, PSL University, CNRS, Sorbonne Universit\'e, Universit\'e de Paris, 5 place Jules Janssen, 92195 Meudon, France
           \and
           LMU M\"{u}nchen, Universit\"{a}tssternwarte, Scheinerstr. 1, 81679, M\"{u}nchen, Germany
           \and
           Astronomical Observatory, University of Warsaw, Aleje Ujazdowskie 4, 00-478 Warszawa, Poland
}

   \date{Received: 11 November 2019; accepted: 18 February 2020}

  \abstract{
  Despite their rarity, massive stars dominate the ecology of galaxies via their strong, radiatively-driven winds throughout their lives and as supernovae in their deaths. However, their evolution and subsequent impact on their environment can be significantly affected by the presence of a magnetic field. While recent studies indicate that about 7\,\% of OB stars in the Milky Way host strong, stable, organised (fossil) magnetic fields at their surfaces, little is known about the fields of very massive stars, nor the magnetic properties of stars outside our Galaxy. We aim to continue searching for strong magnetic fields in a diverse set of massive and very massive stars (VMS) in the Large and Small Magellanic Clouds (LMC/SMC), and we evaluate the overall capability of FORS2 to usefully search for and detect stellar magnetic fields in extra-galactic environments.  We have obtained FORS2 spectropolarimetry of a sample of 41 stars, which principally consist of spectral types B, O, Of/WN, WNh, and classical WR stars in the LMC and SMC. Four of our targets are Of?p stars; one of them was just recently discovered. Each spectrum was analysed to infer the longitudinal magnetic field. No magnetic fields were formally detected in our study, although Bayesian statistical considerations suggest that the Of?p star SMC~159-2 is magnetic with a dipolar field of the order of 2.4 to 4.4\,kG. In addition, our first constraints of magnetic fields in VMS provide interesting insights into the formation of the most massive stars in the Universe.  
    }
    
\titlerunning{A search for magnetic fields in stars in the Magellanic Clouds}
\authorrunning{S. Bagnulo, G.A. Wade, Y. Naz\'e et al.}
   \maketitle
%

\section{Introduction}

Magnetic fields are found in a wide variety of stars across the Hertzsprung-Russell diagram and in two quite different flavours. In the Sun and essentially all other low-mass stars ($M\ltsim 1.5~{\rm M_\odot}$), vigorous magnetic activity is ubiquitous and responsible for most of the observed variability, and it is also essential to their formation and evolution. The magnetic fields have a complex structure,  which generally changes on a timescale of days, weeks, or months; additionally, they produce non-thermal phenomena, such as a hot corona, prominences, flares, and sunspots. The fields are generally believed to be generated by a dynamo, which is produced by a combination of the motions of the deep convective outer layer of the star, together with the Coriolis force due to rapid rotation, so that their strength often increases with the star's rotational velocity \citep[see, e.g.][]{Reiners12}. 

Despite the fact that stars with $M>1.5~{\rm M_\odot}$ do not host a significant outer convective envelope, some 5-10\,\% of these higher-mass stars host strong magnetic fields \citep[e.g.][]{DonLan09,2017MNRAS.465.2432G,2019MNRAS.483.2300S}. These fields have a much smoother morphology than in solar-type stars, that is, organised on a larger scale, and their structure and geometry do not change on the timescales that have been observed so far \citep[up to several decades; see for instance][]{2012MNRAS.419..959O,2018MNRAS.475.5144S,2019MNRAS.483.3127S}. They do not produce surface stellar activity, and their strength does not increase with increasing stellar rotation rate: In fact, due to magnetic braking, the slowest rotators generally have very strong fields \citep[e.g.][]{2019MNRAS.490..274S}. The favoured hypothesis is that the fields are fossil remnants of the Galactic field that accumulated and possibly enhanced during an early phase of their evolution \citep[e.g.][]{Mestel01,Moss01,Neietal15p}. A competing hypothesis put forward in recent years suggests that the magnetic fields in higher-mass stars are the result of either a stellar merger event or a binary mass-transfer event \citep[e.g.][]{TutFed10,2019Natur.574..211S}. The merger hypothesis \citep[e.g.][]{Feretal09,BraSpr04} is also proposed to account for the highly-magnetic white dwarfs \citep{Touetal08}. This scenario seems to be supported by the low incidence rate of magnetic massive stars in close binaries, but it is severely challenged by the detection of the doubly magnetic system $\epsilon$~Lup \citep{Shuetal15}; furthermore, a study of massive close binaries with past or ongoing interaction did not reveal a larger incidence rate of magnetism \citep{Nazetal17}.

Observations of magnetic fields rely on the analysis of the Zeeman effect on the Stokes profiles of spectral lines (mainly Stokes $V$), which can only be normally detected in the brightest or in the most strongly magnetic stars. In fact, all of our knowledge regarding stellar magnetic fields is based on observations of Galactic stars. Therefore, seeing how the magnetic field acts in different environments with different metallicity would help to understand how magnetic fields, in particular fossil fields, originate, evolve, and interact with the circumstellar environment. 

In that respect, hot, massive stars ($M\gtrsim 8\,{\rm M_\odot}$) are especially interesting targets for two reasons. On the one hand, they are bright enough to be observed at a reasonably high signal-to-noise ratio (\snr) in the galaxies that are closest to ours. On the other hand, their role in the chemistry of the Universe is especially important. They comprise only a small fraction of the stellar population of galaxies, yet they contribute a disproportionate amount of energy, matter, and momentum into their host galaxies throughout their lives via their strong, radiatively-driven winds, and when they die as supernovae \citep[e.g.][]{Croetal10}.

Knowledge of the magnetic properties of Galactic O and B stars has advanced remarkably over the last decade, largely due to the Magnetism in Massive Stars \citep[MiMeS;][]{Wadetal16} and B-Fields in OB stars \citep[BOB;][]{Fosetal15} projects. These large-scale surveys have established their statistical incidence in the Galaxy \citep[$\sim 7$\,\% of B- and O-type stars are magnetic;][]{2017MNRAS.465.2432G}, the basic characteristics of these fields \citep[i.e. strong, stable, and organised; e.g.][]{2018MNRAS.475.5144S}, and they have begun to explore the rotational and magnetic evolution of these populations \citep[e.g.][]{2019MNRAS.490..274S,2019MNRAS.489.5669P}. Furthermore, we are able to observe and model the important dynamical effects of the magnetic field on their stellar wind \citep[e.g.][]{Gruetal12,ud-etal13,Nazetal14}, and we are able to model and measure the magnetic spindown \citep[e.g.][]{2009MNRAS.392.1022U,Towetal10}. Recent theoretical studies  \citep[e.g.][]{Meyetal11,Kesetal17,Kesetal19} have also established basic predictions regarding the impact of magnetic fields on the lifetime and internal processes of these stars and, in turn, their impact on the late stages of stellar evolution and the properties of their degenerate remnants \citep[e.g.][]{Petetal17,Geoetal17}. 

An obvious next step towards improving our knowledge on the formation and characteristics of these fields is their study in extra-galactic environments. In fact, candidate extra-galactic magnetic hot, massive stars have recently been identified: a sample of five Of?p stars in the Large Magellanic Cloud (LMC) and Small Magellanic Cloud (SMC). These targets exhibit the telltale spectral peculiarities of the Of?p class \citep{Walborn72,Walborn73} of which all Galactic examples are known to host strong magnetic fields \citep{2017MNRAS.465.2432G}. In particular, they exhibit the characteristic periodic photometric and spectral modulation expected of hot magnetic stars \citep{Nazetal15,Waletal15}. The first attempts to detect magnetic fields in these targets were carried out by \citet[][hereafter referred to as Paper~I]{Bagetal17a} who obtained FORS2 circular spectropolarimetry of them. No magnetic fields were formally detected with longitudinal field uncertainties as small as 350\,G. In Paper~I, we conclude that the magnetic fields of Of?p stars in the Magellanic Clouds are probably not much stronger, on average, than those of similar stars in our Galaxy.

Systematic observing campaigns, such as MiMeS, have mainly targeted stars with masses $\la 60 {\rm M_\odot}$, and only a handful of observations exist for stars with higher masses. No measurements exist for the stars with $M \ga 100\,{\rm M_\odot}$. These very massive stars (VMS) are objects that are more massive than normal O-type stars and are identified as WNh stars (or Of/WN) stars, which are Wolf-Rayet stars with hydrogen in their spectra, and are thought to still be in the core hydrogen burning main sequence phase. The formation of these stars has been speculated to be the result of stellar mergers, although they may also simply form by disc fragmentation \citep{kru15}. It has also been speculated that stellar mergers may be responsible for the generation of fossil magnetic fields of canonical massive stars \citep{sch16}, a scenario that is backed up by magnetohydrodynamic simulations \citep{2019Natur.574..211S}. If VMS formation is attributable to stellar merging, one might therefore expect to find strong dipolar magnetic fields in VMS. 

Setting constraints on the strength of the magnetic fields in VMS would have far reaching consequences regarding the origin of strong, stable magnetic fields in higher-mass stars. For instance, a magnetic incidence fraction $\ga 10$\,\% among very massive stars would provide strong evidence for the merger hypothesis as the origin of magnetism in higher-mass stars. In addition, spectropolarimetric observations of VMS serve the purpose of testing current dynamo models as alternative origins of magnetism in massive stars. In particular, VMS represent the ideal population of stars to test the sub-surface convection models of \citet{Canetal09}.  As  shown by \citeauthor{CanBra11} (\citeyear{CanBra11}, see their Fig.~1), the surface field strength of the magnetic field that is generated via this process is strongest in stars with very high masses ($\ga 120\,M_\odot$), reaching a minimum strength of $\sim 300$\,G.  Furthermore,  theoretical  models  presented by \citet{Yusetal13} show that the core masses of these VMS approach over 90\,\% of the total stellar mass. MHD simulations indicate that the core dynamo-driven fields of main sequence B-type stars should be exceptionally strong \citep{2016ApJ...829...92A}. Since the cores of VMS are larger, closer to the surface, and more vigorously convective, their putative core magnetic fields may be more easily detectable than at the stellar surface of less massive stars.

The purpose of this study is to extend the investigation carried out in Paper~I to a larger and more diverse sample of stars (notably including VMS), in order to revisit candidate magnetic stars and to establish a better understanding of the realistic prospects for the detection of magnetic fields in B- and O-type stars outside the Milky Way. In order to accomplish this goal, we have obtained FORS2 Stokes $V$ observations of extra-galactic Of?p stars (some have been previously studied, in Paper~I, and another that had never been observed with spectropolarimetry before), seven extra-galactic VMS (with $M\ga 100\,{\rm M_\odot}$), as well as a heterogenous sample of 35 (extra-galactic) stars located within a radius of 2-3 arcmin from the main targets, which were observed simply to take advantage of the instrument's multi-object capabilities. 

This paper is organised as follows:
In Sect.~\ref{Sect_Observations} we describe our observing strategy and summarise the technique for data reduction and magnetic field measurements, which are reported in Sect.~\ref{Sect_Results}. 
In Sect.~\ref{Sect_Discussion} we apply some statistical considerations and derive range estimates (almost exclusively upper limits) for the dipolar field strengths of the magnetic poles of the observed stars, and we discuss the various classes of star that we have observed: Of?p stars (Sect.~\ref{Sect_O}), main sequence and evolved OBA stars (Sect.~\ref{Sect_OBA}), cool supergiants (Sect.~\ref{Sect_SG}), and WR stars (Sect.~\ref{Sect_WR}). Implications for magnetic wind confinement in the circumstellar environments of VMS are developed in Sect.\ \ref{subsect:vms_eta}. In Sect.~\ref{Sect_Conclusions} we present our conclusions.

\section{Observations}\label{Sect_Observations} 
To search for magnetic fields in our targets, we used the FORS2 instrument \citep{AppRup92,Appetal98} on the ESO VLT. FORS2 is a multipurpose instrument that is capable of imaging and low resolution spectroscopy, and it is equipped with polarimetric optics (a retarder waveplate and a Wollaston prism). Our observations were obtained in the context of two observing programs, one aimed to continue the spectro-polarimetric monitoring of Of?p stars located in the LMC and SMC (see Paper~I), and one aimed at searching for magnetic fields in the most massive stars.

With very few exceptions, observations were obtained in polarised multi-object spectroscopy (PMOS) modes, using grism 1200B. FORS2 MOS employs a system of mechanical slitlets that can be translated along horizontal tracks. In PMOS mode, nine independent slitlets are available over a $6.8\arcmin \times 6.8\arcmin$ field of view. In addition, the instrument can be rotated, providing two degrees of freedom to position the slitlets. Typically, the observing procedure for PMOS is to fill as many slitlets as possible with primary targets, which are constrained by the position of the targets in the field, and then to fill the remainder with secondary targets with brightnesses comparable to those of the primary targets. As a consequence, we obtained a mix of observations of primary targets and other stars in the field. The actual spectral range depends on the position of the MOS slitlet in the field of view; with the slit in a central position, it was 3700--5120\,\AA.

 The observing programme on extra-galactic Of?p stars was carried out in visitor mode from 20 to 24 November 2017 (programme ID 100.D-0670). The primary targets of this programme were the Of?p stars SMC\,159-2 and AzV\,220, which were already observed in Paper~I, and the newly discovered extra-galactic Of?p star LMCe\,136-1 \citep{Neuetal18}. Most of these observations were carried out in multi-object spectropolarimetric mode with grism 1200B; however, during one night, the star SMC\,159-2 was observed in single object ('fast') mode. In addition, spectroscopic-mode (long-slit) observations of these Of?p stars  were obtained with grism 1200R to monitor the variability of their H$\alpha$ line. Since our hot and weakly reddened targets emit more flux in the blue than in the red and since the blue spectral region is much richer in lines than the red spectral region, for these observations, we employed the  EEV CCD (previously used in the now decommisionned FORS1 instrument), which is optimised in the blue.  All the observed targets are summarised in Table~1.

Observations of reference magnetic stars are not included in the standard FORS2 calibration plan; nevertheless, they are needed to confirm that the position angle of the retarder waveplate is correctly reported by the instrument encoders. For this reason, we decided to use some of the twilight time to observe two well-known and bright magnetic Ap stars: HD\,94660, which has an almost constant longitudinal magnetic field of $-$2\,kG \citep[e.g.][]{Lanetal14}, and HD\,188041, which has a well-known magnetic curve that varies with a period of 223.78\,d \citep{LanMat00} and has been observed for more than 60 years, beginning with \citet{Babcock54}.

A second observing programme (programme ID 094.D-0533) was aimed to study the magnetic properties of the most massive known stars. This programme focused on the 30 Dor region in the LMC, a region that has also been extensively studied as part of the VLT-FLAMES Tarantula Survey \citep[VFTS;][]{Evetal11}. The EEV CCD is not offered in service mode; therefore, we used the MIT CCD, which is optimised for the red.
Our main targets were Mk\,25 and Mk\,51 in NGC 2070 \citep[according to][]{Melnick85}, R136b and c in R136 \citep{Feietal80,Schetal09} and others situated further out \citep[e.g. VFTS\,682;][]{Bestenlehner682}, and NGC 3603, which hosts several well-known massive stars \citep[e.g. NGC 3603a1, B and C;][]{AFJM04,M08,Croetal10}. 
Ultimately only a sub-sample of the original targets were observed: 
VFTS\,621 \citep[in 'Knot 2' from][]{WalBla87}  = Walborn 2 ($M=104\,{\rm M_\odot}$),
VFTS\,506 = Mk\,25    \citep[$M=138\,{\rm M_\odot}$, both mass values from][]{Sabetal14},
R136c = VFTS\,1025    ($M=132\,{\rm M_\odot}$),
Mk37Wa = VFTS\,1021   ($M=141\,{\rm M_\odot}$),
VFTS\,457 = Mk\,51    ($M=100\,{\rm M_\odot}$),
VFTS\,682             ($M=153\,{\rm M_\odot}$), and
Mk\,42 = BAT\,99 105 (= Brey\,77) \citep[$M=153\,{\rm M_\odot}$, all five values from][]{Besetal14}.
In addition to these primary targets, as explained before, nearby stars were used to fill the unoccupied MOS slitlets. These secondary targets were mostly O-type and WR stars, along with a few later-type supergiants. The targets observed in the context of this programme are summarised in Table~3.

The photometric data used in this analysis were obtained by the Optical Gravitational Lensing Experiment (OGLE) project, which was realised on the 1.3\,m Warsaw Telescope located in Las Campanas Observatory, Chile, and operated by the Carnegie Institution for Science during its second, third, and fourth phases \citep[1997-2019,][]{Udaetal15}.

\subsection{Data reduction, magnetic field diagnosis, and quality checks}
Data were reduced using the method explained by \citet{Bagetal15}. The mean longitudinal magnetic field \bz\ (i.e. the component of the magnetic field averaged over the visible stellar disc)
was calculated by minimising the expression
\begin{equation}
\chi^2 = \sum_i \frac{(y_i - \bz\,x_i - b)^2}{\sigma^2_i}
\label{Eq_Chi}
\end{equation}
\noindent where, for each spectral point $i$, 
$y_i$ is the reduced Stokes parameter  $V/I= \pv(\lambda_i)$ and
$x_i = -g_\mathrm{eff} \cz \lambda^2_i (1/I\ \, \mathrm{d}I/\mathrm{d}\lambda)_i$; 
$g_\mathrm{eff}$ is the effective Land\'e factor, \cz\ is a constant 
$\simeq 4.67\,10^{-13}$\,\AA$^{-1}\,{\rm G}^{-1}$ \citep[see][]{Bagetal02}, $\lambda$ is the
wavelength measured in \AA, and $b$ is a free parameter introduced to account for possible spurious polarisation in the continuum. As a quality check, field measurements were also estimated from the null profiles \citep[see][for an extensive discussion on the use of null profiles for quality control]{Bagetal12}. 

For the field measurement, we considered three cases: In Eq.~(\ref{Eq_Chi}) we first used the spectral points and only included H Balmer lines \citep[adopting $\geff=1$ for the Land\'e factor;][]{CasLan94}, then we only included He and metal lines (setting for metal lines on the average value of \geff=1.25),
and finally we included all spectral lines together (H, He, and metal). However, if lines showed obvious emission, we avoided them so as to probe the stellar photosphere rather than the circumstellar environment, unless only emission lines were present in the stellar spectrum, which is similar to the case of WR stars. As the results of these three measurement procedures roughly agree, here, we only report the last value, which also yields the smallest error bars. 
We also measured the so-called null field \nz\  by minimising the expression of the $\chi^2$ of Eq.~(\ref{Eq_Chi}) using the null profiles \nnv\ \citep{Donetal97,Bagetal09} instead of the reduced Stokes profiles \pv. The null profiles are essentially defined as the difference between reduced Stokes parameters obtained from different pairs of measurements, and they represent an experimental estimate of the noise. We expect null profiles and null field values to be consistent with zero within their photon-noise uncertainty. A deviation from zero reveals the presence of significant non-photon noise, while consistency with zero does not guarantee that systematic errors are still absent from the data. For a definition and full discussion of the use of the null profiles and null fields as a quality check, see \citet{Bagetal12} and \citet{Bagetal13}.

\begin{table}
\caption{\label{Tab_Log_EW} New EW measurements obtained with grisms 1200B and 1200R. Uncertainties were estimated from the standard deviation of the measurements of multiple frames (hence they are not reported when obtained from a single frame only).}
{\small
 \begin{tabular}{lllcr@{$\pm$}l}
 \hline \hline
 STAR      & DATE      &  UT &  Line  &\multicolumn{2}{c}{EW (${\rm \AA}$)} \\
             \hline
 SMC\,159-2& 2017-11-21&02:53& He\,{\sc ii} 4686 &$-3.65 $& 0.09 \\
           &           &     & H$\beta$  &$-1.82 $& 0.09 \\
           & 2017-11-21&00:32& H$\alpha$ &\multicolumn{2}{c}{$-11.02$}  \\[1mm]
           & 2017-11-22&01:28& He\,{\sc ii} 4686 &$-4.21 $& 0.08 \\
           &           &     & H$\beta$  &$-2.29 $& 0.07 \\
           & 2017-11-22&00:20& H$\alpha$ &\multicolumn{2}{c}{$-15.50$}  \\[1mm]
           & 2017-11-24&02:25& He\,{\sc ii} 4686 &$-4.76 $& 0.05 \\
           &           &     & H$\beta$  &$-2.84 $& 0.06 \\
           & 2017-11-24&00:53& H$\alpha$ &$-19.72$& 0.17 \\[2mm]
LMCe\,136-1& 2017-11-21&06:40& He\,{\sc ii} 4686 &$-2.36 $& 0.03 \\
           &           &     & H$\beta$  &$-1.09 $& 0.03 \\
           & 2017-11-21&08:24& H$\alpha$ &$-7.72 $& 0.12 \\[1mm]
           & 2017-11-22&08:20& He\,{\sc ii} 4686 &\multicolumn{2}{c}{$-2.46 $}  \\
           &           &     & H$\beta$  &\multicolumn{2}{c}{$-1.36 $}  \\
           & 2017-11-22&08:31& H$\alpha$ &\multicolumn{2}{c}{$-7.66 $}  \\[1mm]
           & 2017-11-23&07:39& He\,{\sc ii} 4686 &$-2.44 $& 0.04 \\
           &           &     & H$\beta$  &$-1.54 $& 0.03 \\
           & 2017-11-23&08:43& H$\alpha$ &\multicolumn{2}{c}{$-8.70 $}  \\[2mm]
 LMC\,164-2& 2017-11-22&07:51& He\,{\sc ii} 4686 &\multicolumn{2}{c}{$-1.47 $}  \\
           &          &     &  H$\beta$  &\multicolumn{2}{c}{$-0.47 $}  \\
           &2017-11-22&08:31&  H$\alpha$ &\multicolumn{2}{c}{$-3.25 $}  \\[2mm]
 AzV 220   &2017-11-24&04:51&  He\,{\sc ii} 4686 &$-1.33 $& 0.02 \\
           &          &     &  H$\beta$  &$-1.94 $& 0.04 \\
 \hline
 \end{tabular}
 }
 \end{table}
 
\begin{figure}
\begin{center}
\scalebox{0.42}{
\includegraphics*[trim={0.8cm 4.7cm 0.3cm 2.8cm},clip]{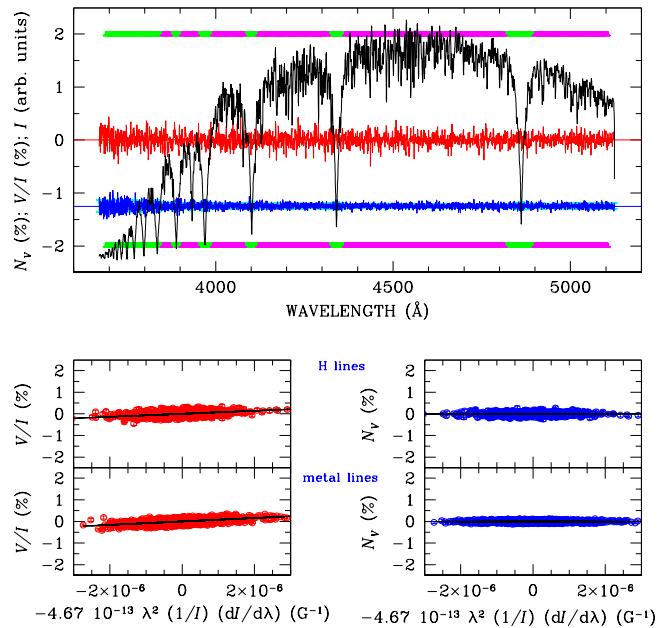}}
\caption{\label{Fig_HD188041a} 
  FORS2 observations of the magnetic reference Ap star HD\,188041.  In the upper panel, the black solid line shows the Stokes $I$ spectrum (uncorrected for the transmission function of the atmosphere + telescope and instrument optics); the red solid line shows the reduced Stokes $V$ spectrum, $\pv=V/I$ (in \% units), and the blue solid line is the null profile offset by $-1.25$\,\% for display purposes. The scattering of the null profile about zero is consistent with the 1\,$\sigma$ photon-noise uncertainties, which are also shown centred around  $-1.25$\,\% and appear as a light blue background to the null profile.  Spectral regions highlighted by green bars (at the top and at the bottom of the panel) have been used to determine the \bz\ value from H Balmer lines, while the magenta bars highlight the spectral regions used to estimate the magnetic field from He and metal lines. The four bottom panels show the best-fit obtained by minimising the $\chi^2$ expression of Eq.~(\ref{Eq_Chi}) using the \pv\ spectra (left panels) and the \nnv\ spectra (right panels) for H Balmer lines (upper panels) and metal lines (lower panels). 
  }
\end{center}
\end{figure}
\begin{figure}
\begin{center}
\scalebox{0.40}{
\includegraphics*[trim={0.8cm 5.0cm 0.3cm 3.0cm},clip]{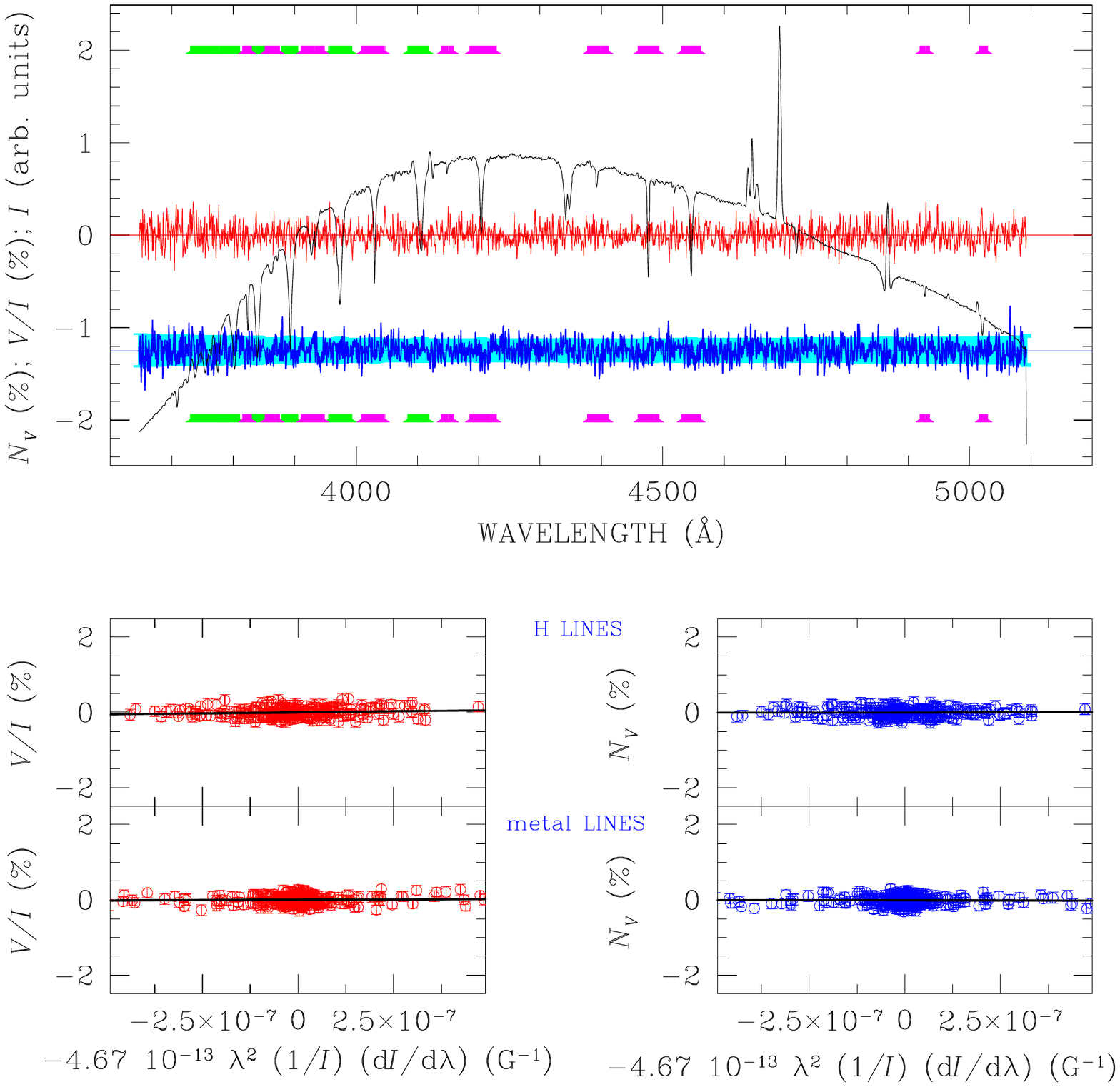}}
\scalebox{0.40}{
\includegraphics*[trim={0.8cm 5.0cm 0.3cm 3.0cm},clip]{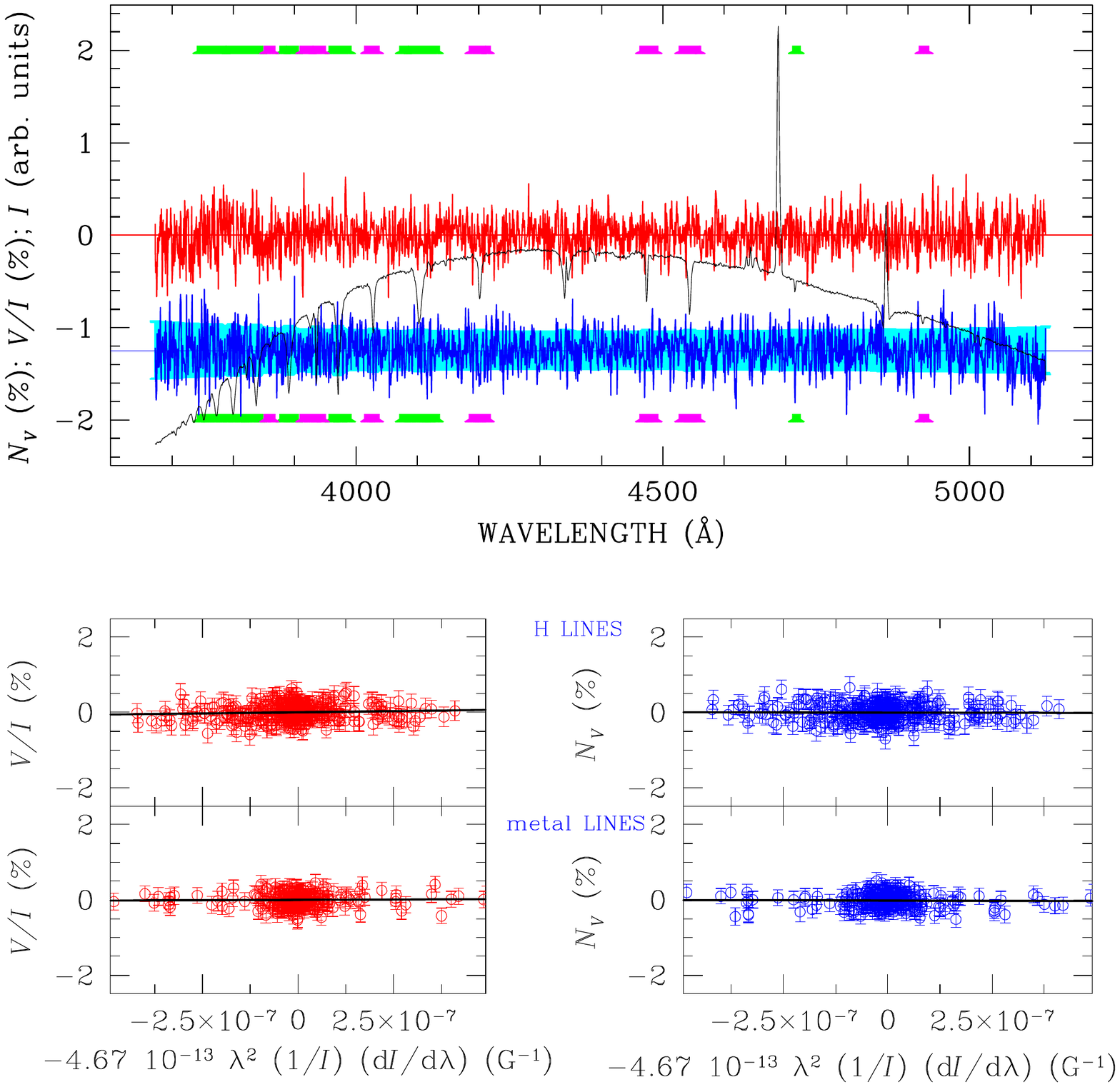}}
\scalebox{0.40}{
\includegraphics*[trim={0.8cm 5.3cm 0.7cm 3.0cm},clip]{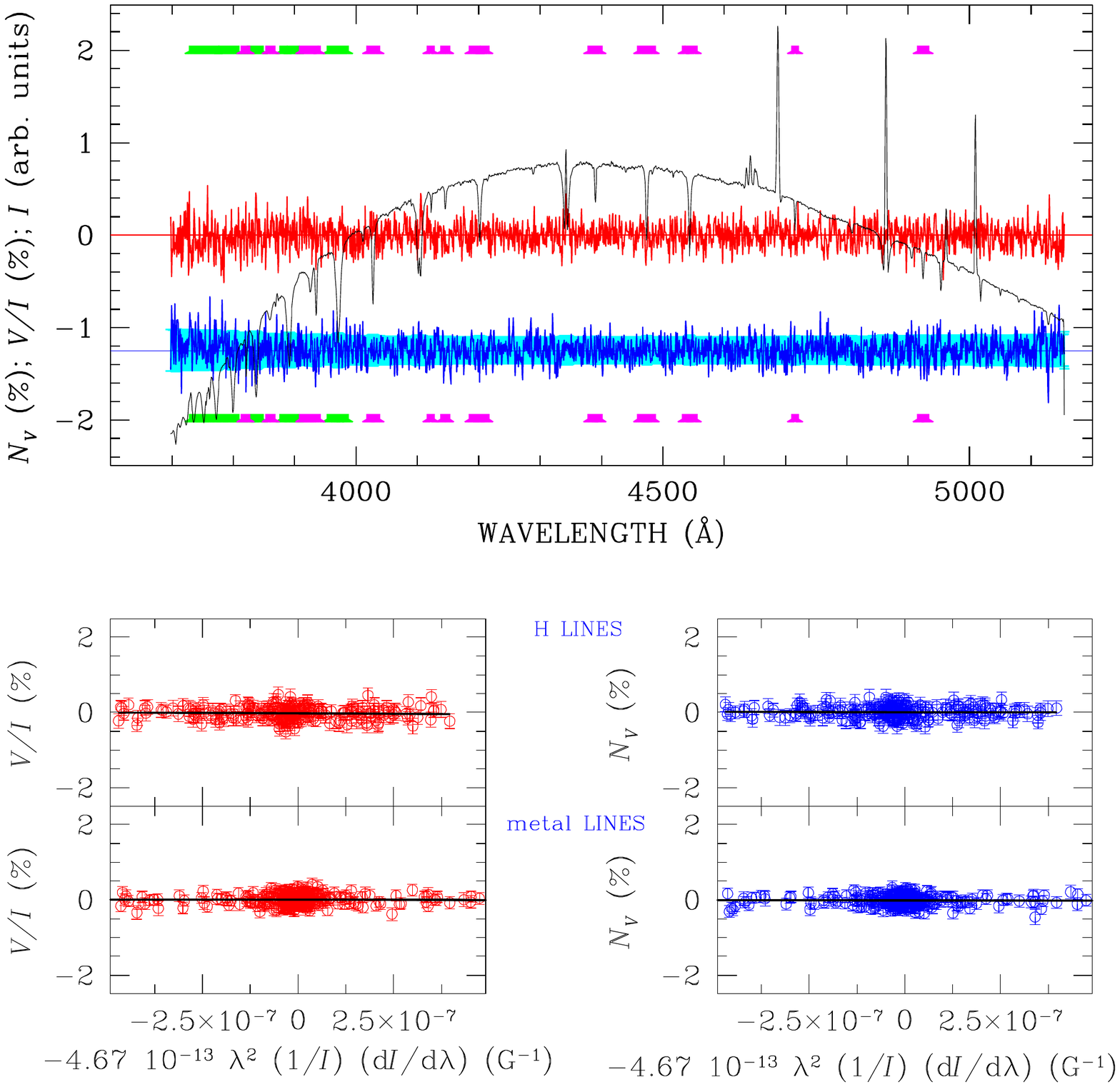}}
\caption{\label{Fig_OStars} Same as Fig.~\ref{Fig_HD188041a} for
  three of our science targets: LMCe\,136-1 (top panels),
SMC\,159-2 (mid-panels)
and AzV\,220 (bottom panels).
}
\end{center}
\end{figure}

\section{Results}\label{Sect_Results}
\begin{table}
  \caption{\label{Table_Limits} Limits for the dipolar field strength obtained through Bayesian statistical considerations.
  Column 2 reports the \Bupp\ value, and the flag of col.~3 means no (N), weak (w), moderate (m),
  strong (s), or very strong (vs) evidence for the presence of a magnetic field. 
  Columns 4 and 5 as cols.~2 and 3, but for the null field values. The meaning of relative and absolute refers to results obtained assuming that the star's rotation period is known, as explained in the text.}
\begin{tabular}{lrcrc}
\hline\hline
STAR       & \multicolumn{2}{c}{\Bupp} & \multicolumn{2}{c}{\Nupp} \\
\hline
 & & & & \\[-2mm]
{\bf Of?p:}                &       &   &      &  \\
AzV\,220                   &  2325 & N & 8825 & N\\
SMC~159-2                  & 12125 & m &10725 & N\\
SMC~159-2 (relative)       & 10075 & s & 7275 & N\\
SMC~159-2 (absolute)       & 11275 &vs & 9825 & N\\
LMCe\,136-1                &  2825 & N & 1375 & N\\
LMCe\,136-1 (relative)     &  3225 & N & 1475 & N\\
LMCe\,136-1 (absolute)     &  4625 & N & 2275 & N\\
LMC\,164-2                 &  3775 & N & 7975 & N\\
LMC\,164-2 (relative)      &  3775 & N & 7975 & N\\
LMC\,164-2 (absolute)      &  6725 & N & 2525 & N\\
\hline
 & & & & \\[-2mm]
{\bf OB supergiants:}                 &       &   &      &  \\                       
Mk\,42     (VMS)           &  3725 & N & 1825 & N\\      
VFTS\,526                  &  7975 & N & 5725 & N\\      
Mk37Wa     (VMS)           &  6925 & N & 8825 & N\\[1mm] 
VFTS\,291                  &  4275 & N & 6175 & N\\      
VFTS\,458                  &  2075 & N & 3525 & N\\      
\hline
 & & & & \\[-2mm]
{\bf BA giants:}        &       &   &      &  \\                       
NGC\,346 ELS\,1080         &  9475 & N & 5375 & N\\     
NGC\,346 ELS\,19           &  7025 & N & 6075 & N\\ 
\hline
 & & & & \\[-2mm]
{\bf OB dwarfs:}        &       &   &      &  \\                       
VFTS\,441                  &  5875 & N & 4875 & N\\
VFTS\,500                  &  7075 & N &18925 & w\\
VFTS\,506  (VMS)           &  2474 & N & 3575 & N \\
VFTS\,586                  & 12285 & N & 5565 & N\\ 
VFTS\,621  (VMS)           &  1525 & N & 1275 & N\\ [1mm]
VFTS\,589                  &  9025 & N & 7225 & N\\      
AzV\,55                    &  1025 & N &  975 & N\\      
AzV\,66                    &  1325 & N & 2075 & N\\      
2dFS 5037                  &  8925 & N & 4375 & N\\      
2dFS 5038                  &  1675 & N & 1225 & N\\      
NGC\,346 ELS\,27           & 12975 & N & 4475 & N\\      
NGC\,346 ELS\,68           &  6525 & N & 5525 & N\\      
NGC\,346 ELS\,100          &  5075 & N & 8625 & N\\      
NGC\,346 ELS 103           &  3275 & N &16675 & N\\ [1mm]
\hline
 & & & & \\[-2mm]
{\bf Cool supergiants:}    &       &   &      &  \\
W60 D24                    &   175 & N &  225 & N\\
RM\,1-546                  &   775 & N &  775 & N\\
SkKM\,179                  &  3025 & N &  575 & N\\
VFTS\,341                  &  1075 & N &  625 & N\\
CPD$-$69\,463              &  1175 & N &  775 & N\\
    \hline
 & & & & \\[-2mm]
{\bf WR:}                  &       &   &      &  \\
VFTS\,507                  &   925 & N & 1425 & N\\
VFTS\,509                  &  1175 & N & 1825 & N\\
R\,136c    (VMS)           & 12675 & N & 5575 & N \\
VFTS\,682  (VMS)           &  2975 & N & 3075 & N \\[2mm]
\hline
 & & & & \\[-2mm]
\hline
\end{tabular}
\end{table}

In total, we observed 41 science targets (some of them multiple times), including Of?p stars, cool supergiants, main sequence OB stars, evolved OB stars, and WR stars. The results for each of these groups are discussed in Sects.~\ref{Sect_O} to \ref{Sect_WR}. In addition, two well-known magnetic Ap stars were also observed to check the correct alignment of the polarimetric optics, namely HD\,94660 and HD\,188041. Figure~\ref{Fig_HD188041a} shows an example of field detection on one of the magnetic reference stars (HD\,188041), and Fig.~\ref{Fig_OStars} shows the same plots for the Of?p targets LMCe\,136-1, SMC\,159-2, and AzV\,220.  Our full list of magnetic measurements is given in Tables~\ref{Tab_OP} and \ref{Tab_VMS}. 

In addition to the magnetic field measurements, for the Of?p stars of our target list, we also measured the equivalent widths (EWs) of He~{\sc ii} 4686, H$\beta,$ and H$\alpha$ (see Table~\ref{Tab_Log_EW} and Fig.~\ref{Fig_EWs}); these measurements are discussed in Sect.~\ref{Sect_O}.

\subsection{Observations of magnetic reference stars}
Our field measurement of HD\,94660 ($\bz = -2345 \pm 45$\,G) is consistent with the expected value from previous measurements published in the literature \citep[the star exhibits a longitudinal field that oscillates around $-$2\,kG, see][and references therein]{Bagetal12}, and so were the field measurements of HD\,188041 ($\bz = 770 \pm 20$\,G on the nights of 20 to 21 November 2017 and $780 \pm 20$\,G on the nights of 21 to 22 November 2017). Phased with the ephemeris of \citet{LanMat00}, our measurements of  HD\,188041 appear to have been taken close to the magnetic minimum and, similar to what was found in Paper~I, they are $\sim\,200$\,G higher compared to the measurements obtained by \citet{Babcock54} and \citet{Babcock58} at similar rotation phases. These discrepancies do not suggest a problem with the instrument, but, as discussed in detail by \citet{Lanetal14}, they are rather a symptom of the systematic differences that are present when the longitudinal field is measured with different instruments and even simply different setups.

\subsection{Magnetic field measurements of the science targets}\label{Sect_Field_Meas}
\begin{figure}
\includegraphics*[width=9.0cm,trim={0.8cm 5.0cm 0.3cm 2.5cm},clip]{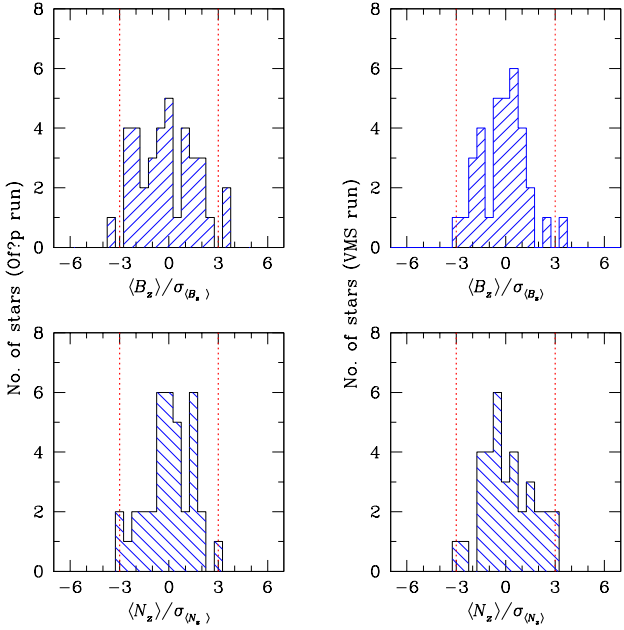}
\caption{\label{Fig_Histos} Distribution of \bz\ and \nz\ values normalised to their uncertainties for the stars observed during the run dedicated to Of?p stars (left panels) and for the stars observed during the run dedicated to the VMS stars (right panels). The vertical dotted lines mark the $3-\sigma$ limits of these distributions.}
\end{figure}

Our field measurements summarised in Tables~\ref{Tab_OP} and \ref{Tab_VMS} mainly represent non-detections, but there are some occasional $\sim 3\,\sigma$ detections, both from the reduced Stokes $V$ profiles and from the null profiles. As discussed by \citet{Bagetal12}, spectropolarimetry with Cassegrain-focus mounted instruments is less accurate than what can be obtained with a fibre-fed instrument that is specifically designed for accurate radial velocity measurements (e.g. ESPaDOnS at the CHFT and HARPS at the 3.6\,m telescope of the La Silla Observatory). The reason is that even tiny instrument flexures occurring over the course of an exposure series may be responsible for spurious signals that mimic a Zeeman signature, especially on relatively narrow spectral lines \citep[see Fig.~1 of][]{Bagetal13}. The situation is even more likely to occur during particularly long exposures at high airmass, such as those obtained during the run dedicated to the Of?p stars. Spurious results may also occur more frequently in the presence of blending, such as in the case of many of those observed in the run dedicated to the very massive stars. The reason being is that seeing variations may change the appearance of spectral features during the observing series. Therefore, the occurrence of marginal detections on null profiles is not particularly surprising. For the same reason, our marginal \bz\ detections on the reduced $V$ profiles do not necessarily suggest the presence of a magnetic field. An overall view of the distribution of our \nz\ and \bz\ measurements, which are normalised by their uncertainties, is given in Fig~\ref{Fig_Histos}. In the ideal case, the distribution of the quantities $\nz\ / \sigma_{\nz}$ should be similar to a Gaussian distribution with $\sigma = 1$. In fact, the  $\nz\ / \sigma_{\nz}$ histograms display a larger width, with some \nz\ detections at $3\,\sigma$ level. This suggests that also the various marginal ($\sim 3\,\sigma$) \bz\ detections should be treated with caution and they should not be considered to be conclusive. In conclusion, the uncertainties declared in our result tables do not fully account for non-photon noise, which although apparent in the global analysis of all measurements, is difficult to quantify for individual stars.

\begin{figure}
\begin{center}
\scalebox{0.45}{
\includegraphics*[trim={0.8cm 1.7cm 0.3cm 0.8cm},clip]{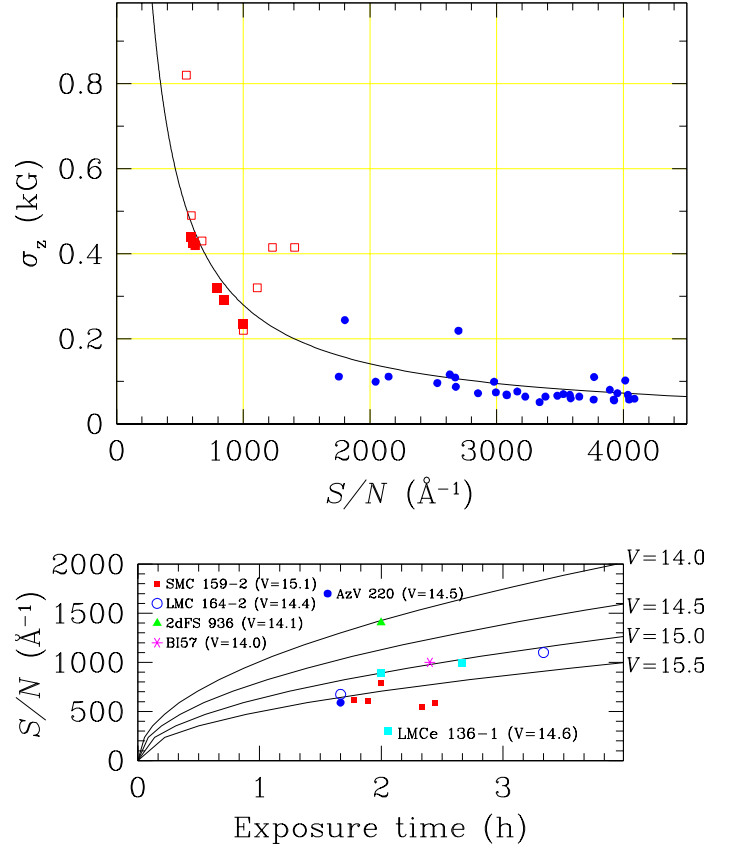}}
\caption{\label{Fig_SNR} {\it Top panel:} uncertainties of the field measurements obtained with the FORS1/2 instruments for various O-type stars. Blue filled circles refer to bright Galactic O-type stars observed with FORS1 \citep[see][]{Bagetal15}; red empty squares correspond to LMC/SMC O-type stars observed with FORS2 in Paper~I and red filled squares LMC/SMC O-type stars observed with FORS2 and presented in this paper. The solid line is the best hyperbolic fit to the data points. {\it Bottom panel:} {\it S/N} vs. exposure time. Symbols
refer to different observations of O-type stars as explained in the legend. Solid lines show the predictions of ETC version P103.2 for a O5 star observed at airmass=1.5 with grism 1200B\ through a 1\arcsec\ slit width, with 1.0\arcsec seeing, and fraction of lunar illumination = 0.5.}
\end{center}
\end{figure}

\subsection{Uncertainties of the field measurements verus {\it S/N}}
Figure~\ref{Fig_SNR} combines previous results obtained with bright
(Galactic) O-type stars observed with FORS1 \citep[shown in Fig.~5
of][]{Bagetal15} with the data of extra-galactic Of?p stars
presented here and in Paper I. 
The top panel shows the uncertainties of our field measurements versus {\it S/N}; the bottom panel shows the predictions of the FORS2 exposure time calculator (ETC) for the {\it S/N} on $V=14-15.5$ magnitude stars that can be reached as a function of the exposure time as well as the {\it S/N} that was actually reached in our observations. Discrepancies between ETC predictions and real observations are due, in part, to less-than-ideal weather conditions and also to the fact that the ETC is optimistic regarding the instrument performances in polarimetric mode. Nevertheless, detection at a $5\,\sigma$ confidence level of a magnetic field in an O-type star of the MC with a dipolar field $\ga 1500-2000$\,G should be just within the limits of the capabilities of the FORS2 instrument if the star is observed in favourable conditions. Indeed, it should be taken into account that a star could have a very strong surface magnetic field, but, for geometrical reasons, have a zero average longitudinal component at the time of the observation. More generally, a single null measurement of the longitudinal field only provides limited constraints on the star's surface dipole field strength. However, when combined together, several measurements on the same target may provide a useful estimate of the upper limit of the star's dipolar field strength. In the next section, we use a statistical approach that proves to be especially useful when no direct modelling of the data are possible (e.g. due to scarcity of data or to the fact that all field measurements are non-detections).

\section{Constraints on the magnetic field strength of the observed extra-galactic stars}\label{Sect_Discussion}
The problem of how to estimate an upper limit for the field strength from a series of non-detections has been tackled using approaches of both Bayesian (e.g. \citealt{KolBag09}; \citealt{PetWad12}; and \citealt{Asetal14}) and classical (frequentist) statistics \citep[e.g.][]{Neietal15}. Here, we follow a Bayesian approach, extending the use of Eq.~(7) of \citet{KolBag09} to an arbitrary number of field measurements \citep[see also Sect.~3.2 of][]{PetWad12}. We calculate the probability ${\cal P}$ that, given a set of $\bz_j \pm \sigma_j$ measurements, the strength of the dipolar field \Bdip\ is within the range $[B_1,B_2]$. Equation~(7) of \citet{KolBag09} becomes
\begin{equation}
\begin{array}{r@{\;}l}
{\cal P} (B_1 \le \Bdip \le B_2)& \propto \\
  & \int_{\Bdip} \int_i \int_\beta
     p_B(\Bdip) \sin i \sin\beta \prod_j \int_{f_j} \\
  & \exp \left( - \frac {(\bz_j - k(u) \Bdip
    [\cos i \cos\beta + \sin i \sin\beta \cos f_j])^2} {2\sigma_j^2} \right) \\
\end{array}
\label{eq:PrB}
\end{equation}
where $i$, the tilt angle between line of sight and rotation axis, and $\beta$, the angle between rotation axis and magnetic field axis, are assumed to be isotropic; the prior probability densities of phase $f_j$ are constant; for the $\Bdip$ distribution, we assume $p_B$ is a modified Jeffreys prior \citep[e.g.][]{PetWad12}; and $k(u)$ is a constant that depends on the limb-darkening coefficient $u$ (assuming $u=0.5$ we have $k = 0.31$). Practically speaking, for most of the cases of this paper, we are primarily interested in the upper limit \Bupp\ of the dipolar field strength defined such that there is a 95\,\%  probability that the dipolar field \Bdip\ at the magnetic pole of the star is smaller than \Bupp \ if the star is magnetic. This estimate of \Bupp\ alone does not, however, tell us whether the star is likely to be magnetic or not. Instead, the likelihood that a magnetic field is present may be estimated by the odds ratio \citep[see Eq.~3.14 of][]{Gregory05}. Following \citet{PetWad12}, we affirm that there is no evidence for a magnetic field if the odds ratio is $\la 3$ and that there is weak, moderate, strong, or very strong evidence that the star is magnetic if the odds ratio is between 3 and 10, 10 and 30, 30 and 100, and $\ga 100$, respectively. As a quality check, we also applied these concepts of Bayesian statistics to the null field values and derived the corresponding maximum dipolar field strength \Nupp. Similar to what is discussed for the individual \nz\ values by \citet{Bagetal12}, the values obtained from the \nz\ values should not be intepreted as errors on the \Bupp\ values; although from the \nz\ values, we should expect to find no evidence for the presence of a magnetic field.

For three Of?p stars for which we have more than one magnetic field measurement,  we have a good estimate of the rotational period; therefore, the relative phase of each measurement is also known, that is, $f_j=f+\phi_j$ for known $\phi_j$. If we assume that the magnetic field maximum occurs at the same time as the maximum of the light curve \citep[e.g.][]{Munetal20}, we can also deduce the rotational phase at which our field measurements were obtained. If the star's rotation period is known, but not the epoch of the maximum of \bz, the product of $N$ $f_j$-integrals in (\ref{eq:PrB}) is replaced by a single integral over $f$, and if the absolute value of the phase is also known, there is no phase integral. Further analysis of this approach will be provided by Asher \& Bagnulo (in preparation).

\begin{figure*}
\includegraphics*[width=9.4cm,trim={0.8cm 6.0cm 0.3cm 2.5cm},clip]{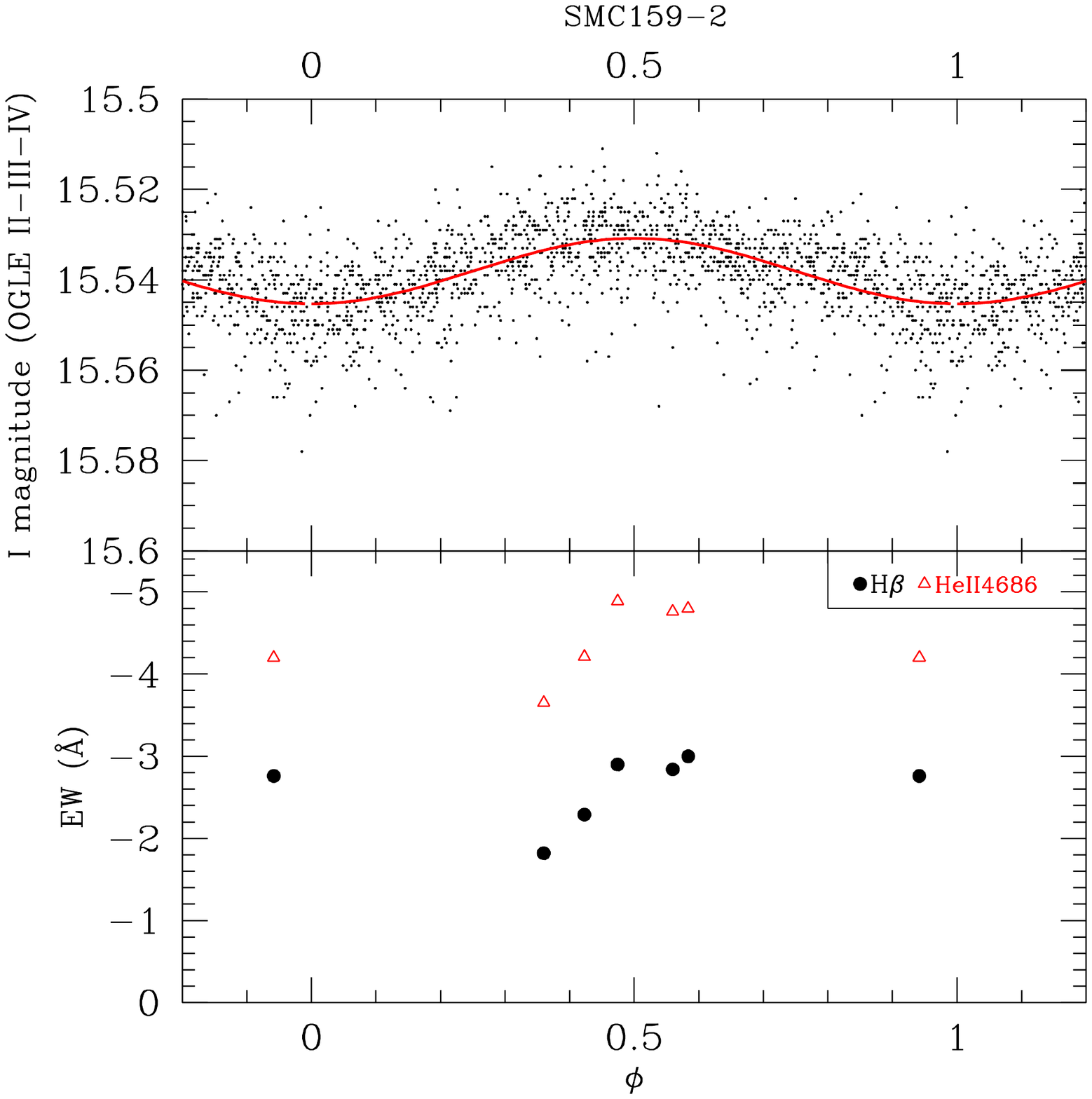}
\includegraphics*[width=9.4cm,trim={0.8cm 6.0cm 0.3cm 2.5cm},clip]{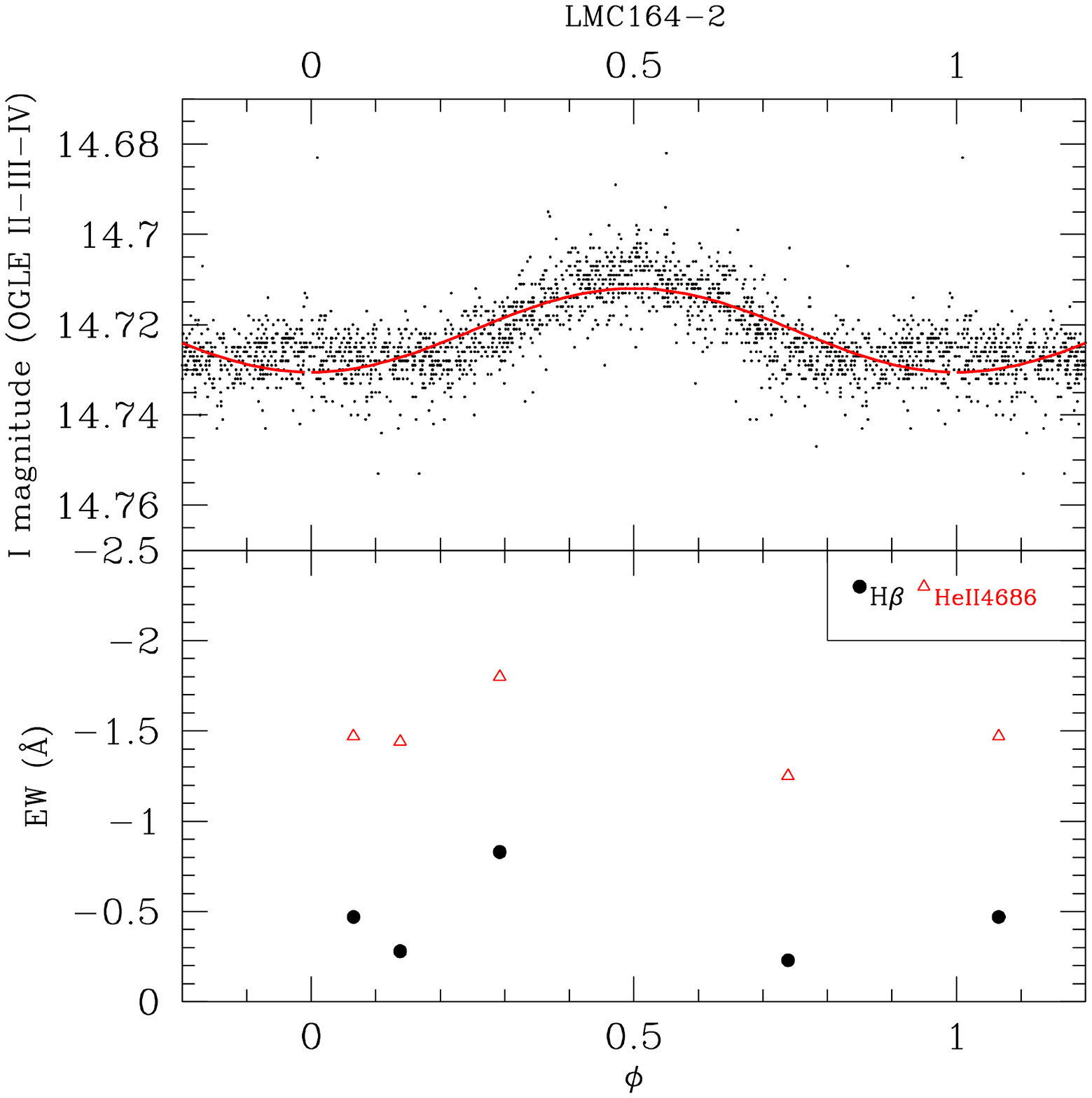}\\
\includegraphics*[width=9.4cm,trim={0.8cm 6.0cm 0.3cm 2.5cm},clip]{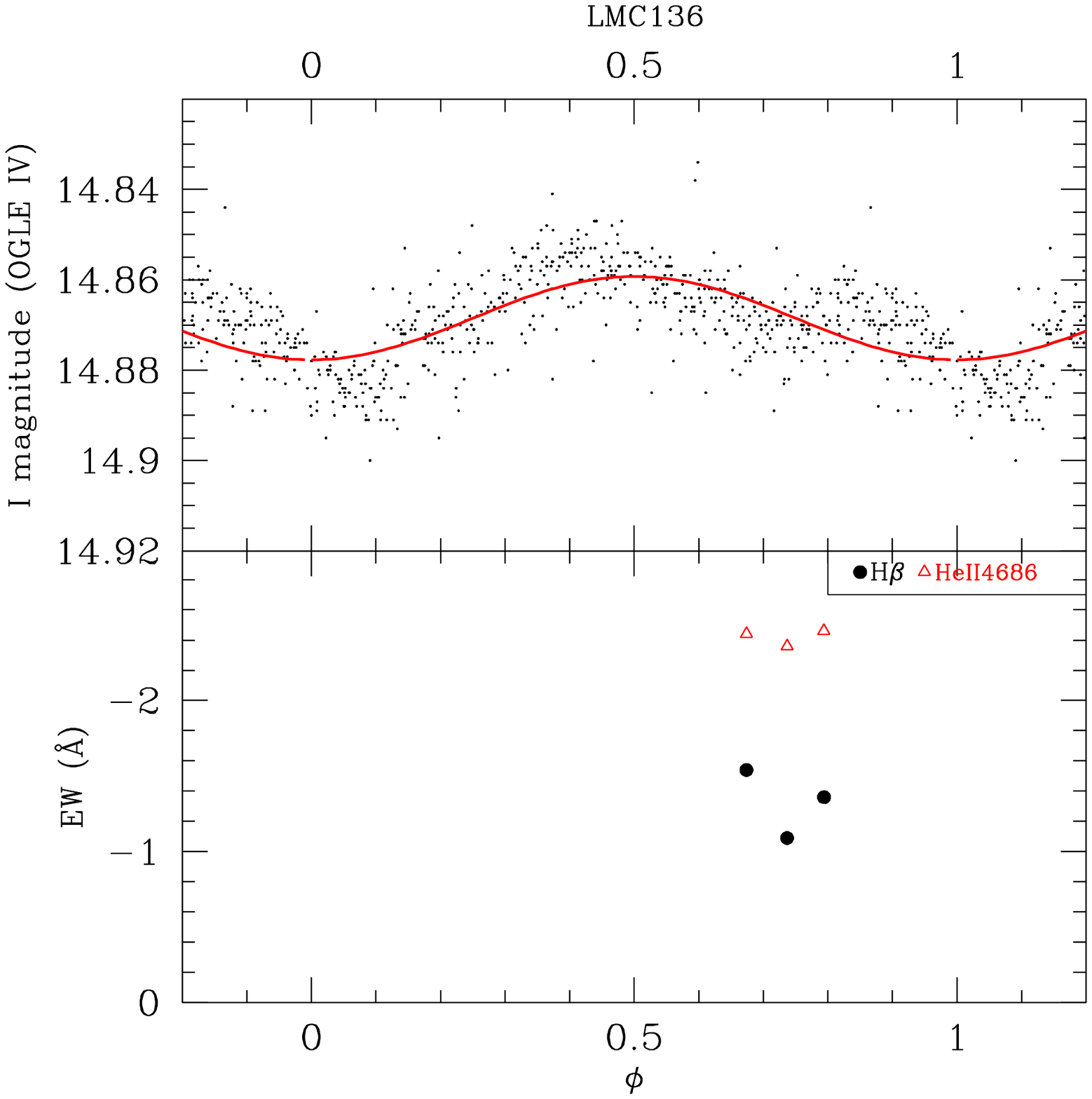}
\includegraphics*[width=9.4cm,trim={0.8cm 6.0cm 0.3cm 2.5cm},clip]{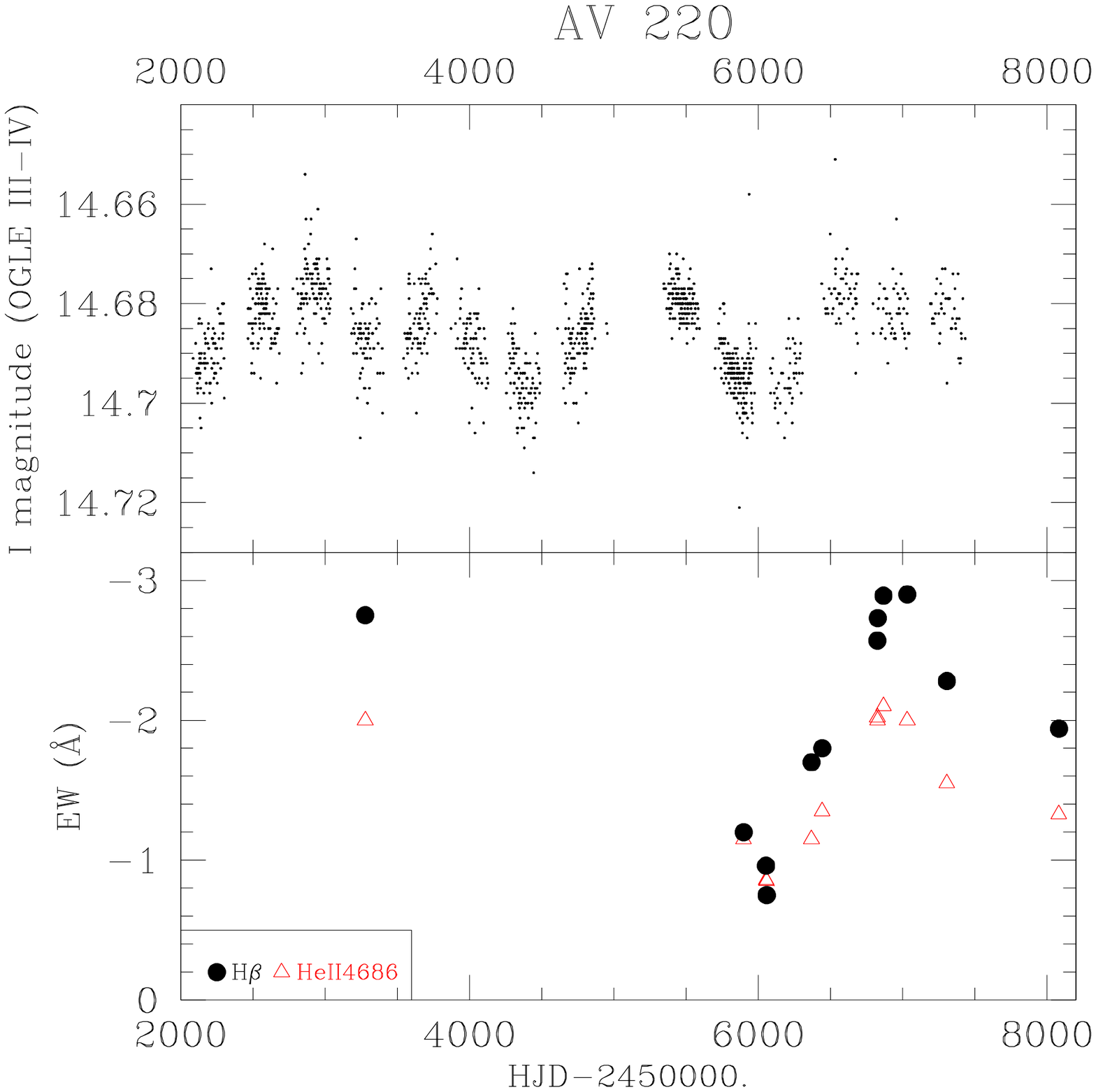}
\caption{\label{Fig_EWs} Comparison of photometric measurements with equivalent widths of four extra-galactic Of?p stars. {\it Top left panel:} SMC\,159-2; {\it top right panel:} LMC\,164-2; {\it bottom left panel:} LMCe\,136-1; {\it bottom right panel:} AzV\,220.  Symbols refer to H$\beta$ (black solid circles) and He~{\sc ii}~$\lambda 4686$ (empty red triangles). Equivalent width data are from Table~\ref{Tab_Log_EW}, from \citet{Bagetal17a}, and the references therein.\ Photometric data are from OGLE.}
\end{figure*}

In the following, we apply this method to the stars of our target list, which include O?fp stars (Sect.~\ref{Sect_O}), early-type, and late-type main sequence and supergiants (Sect.~\ref{Sect_SG}), main sequence OB stars (Sect~\ref{Sect_OBA}), and WR stars (Sect~\ref{Sect_WR}). Our targets range in apparent $V$ magnitude from 11.3 (the red supergiant candidate TYC\,8891-3239-1; Neugent et al. 2012) to 16.2 (luminosity class III-V B-type stars). For the majority of objects, the individual measurements of magnetic fields have uncertainties of a few hundred Gauss; the limitation of the precision is not really due to stellar luminosities but the lack of absorption lines. The estimated upper limits of the dipolar field strengths for all stars observed at least twice (also including the data collected in Paper~I) are given in Table~\ref{Table_Limits}. Generally, these upper limits were estimated assuming that each measurement was obtained at a random value of the rotational phase, which is the only possibility when the star's rotation period is unknown. For three O?fp stars, we could also estimate these upper limits using the constraint given by their rotation period. In this situation we know the phase offset between each field measurement (case 'relative' in Table~\ref{Table_Limits}). Under the hypothesis that the maximum of the light curve coincides with the maximum of the longitudinal magnetic field, we can also make the further assumption of knowing at which position of the magnetic curve each field measurement was obtained (case 'absolute' in Table~\ref{Table_Limits}).

\subsection{Of?p stars}\label{Sect_O}
The Of?p stars of our target list are probably the best magnetic candidates and deserve to be discussed individually.
\subsubsection{SMC\,159-2}
\begin{figure}
\begin{center}
\scalebox{0.45}{
\includegraphics*[trim={0.8cm 4.7cm 0.3cm 2.8cm},clip]{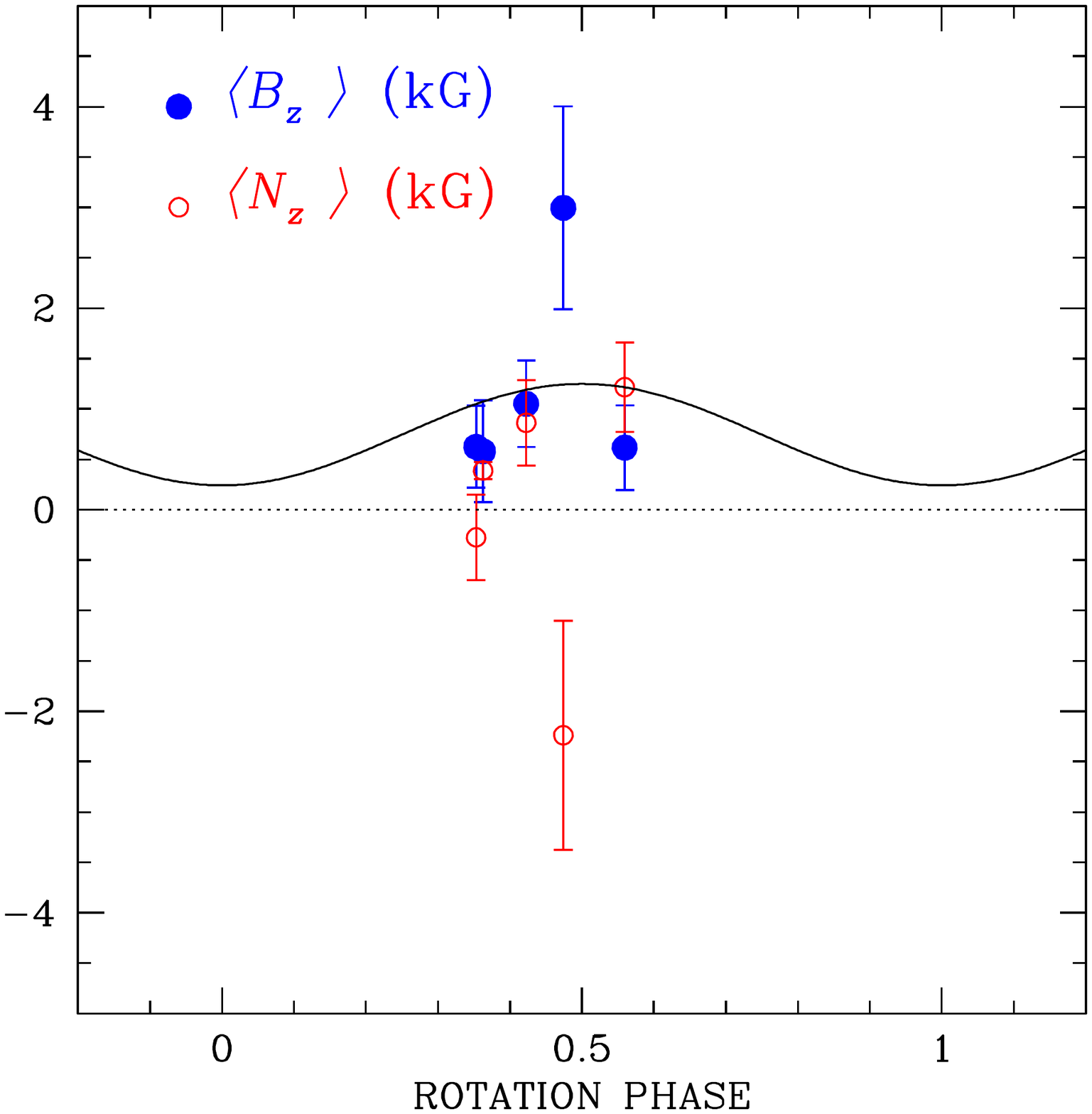}}
\caption{\label{Fig_SMC} Mean longitudinal field and null field measurements of SMC\,159-2 as a function of the rotational phase. Here the rotation phase is defined consistently with the light curves of Fig.~\ref{Fig_EWs}, which is offset by 0.5 cycles from what is usually adopted for longitudinal field curves (which generally have the maximum at rotation phase zero). The field measurement at rotation phase = 0.5 was presented in Paper~I. The solid curve shows the prediction of the model by \citet{Munetal20}.}
\end{center}
\end{figure}
SMC\,159-2 was identified as an Of?p star by \citet{Masetal14}. \citet{Nazetal15} investigated the photometry of this star and establish clear $14.914\pm 0.004$\,d periodic variability. In Paper~I, we demonstrate that the EWs of H$\beta$ and He~{\sc ii}~$\lambda 4686$ varied smoothly according to this same period. Using the most recent epochs of OGLE photometry, we derived a new ephemeris for SMC\,159-2, with $T_0=2457416.997$ and $P=14.915 \pm 0.003\,{\rm d}$. This ephemeris is sufficiently precise to allow us to phase our new EWs, measured from He~{\sc ii} 4686, H$\beta$ and H$\alpha$, with those of \citet{Bagetal17a}. The resultant phase curves, illustrated in Fig.~\ref{Fig_EWs}, show an emission maximum occurring at approximately the same phase as maximum brightness. Such a relative phasing is qualitatively consistent with the rotational modulation of a stellar magnetosphere (see e.g. \citealt{Munetal20} and \citealt{Driessen18}).

In Paper~I, we measured a longitudinal field of $2.8\pm 1$\,kG, which is nearly a 3\,$\sigma$ detection of a 3\,kG longitudinal field. In view of the considerations of Sect.~\ref{Sect_Field_Meas}, this measurement is probably too marginal to be considered to be a reliable detection. In this work, we obtain four new polarimetric measurements (two of which sequentially on a single night), all are close to maximum emission as is the measurement presented in Paper~I. Compared to the observations presented in Paper~I, which were obtained with similar exposure times, our new measurement uncertainties are lower because of much better weather conditions. None of the measurements yield a significant detection of the longitudinal field, with a best uncertainty of 270\,G, but they are all of a positive sign, and roughly consistent with each other (as expected when obtained at  a similar rotation phase if the star is magnetic).  Figure~\ref{Fig_SMC} shows our \bz\ and \nz\ measurements of this star as a function of the star's rotational phase.  

Although none of our measurements may be considered to be a detection, the results of a Bayesian statistical analysis lead to the suggestion that the star possesses a magnetic field. First we consider the case in which we do not use the star's rotation period as a constraint, that is, we assume that all field measurements were obtained at an unknown rotational phase. In this situation, the two consecutive measurements obtained on the night of 21 November 2017 were averaged out and we considered a total of four field measurements. The odds ratio, that is, the probability that the star is magnetic versus the probability that there is no magnetic field is $\sim 13$, which corresponds to 'moderate' evidence for the presence of a magnetic field; the probability distribution function (PDF) of the dipolar field strength is smooth, and it tells us that there is a 50\,\% probability that the dipolar field strength at the pole is between 0 and 5\,kG, or a 95\,\% probability that the dipolar field strength at the pole is between 0 and 12\,kG. If we use the star's rotation period as a constraint, and we assume that the mean longitudinal field has a maximum when the EW has a maximum, then the odds ratio is $\sim 140$, corresponding to very strong evidence for the presence of a magnetic field. This reflects the fact that our five measurements all have the same sign and are all close in the rotation phase (within three days over a 15 day rotation period). On the other hand, there are two negative and three positive \nz\ values in Table~\ref{Tab_OP}. This set of values is consistent with the zero field (thus `no evidence' in Table~\ref{Table_Limits}), while also being consistent with a significant \Nupp value. The \Bdip\ PDF tells us that there is a 50\,\% probability that the dipolar field strength at the pole is comprised between 2.4 and 4.4\,kG, and a 95\,\% probability that is between 0 and 11\,kG. These predictions are consistent with those by \citet{Munetal20}, who used the analytic dynamical magnetosphere (ADM) model of \citet{Owocki16} to compute synthetic lightcurves of four Of?p stars in the Magellanic Clouds, including SMC\,159-2. Comparing their model to the data of SMC\,159-2, \citet{Munetal20} predict a surface dipole field strength of $6.4_{-1.8}^{+3.5}$\,kG.

\subsubsection{LMC\,164-2}
LMC\,164-2 was also identified as an Of?p star by \citet{Masetal14}. \citet{Nazetal15} established clear $7.9606\pm 0.0010$\,d periodic variability. In Paper~I, the EWs of H$\beta$ and He~{\sc ii}~$\lambda 4686$ are shown to be compatible with this period. 

In Paper~I, we measured $\bz=0.20\pm 0.56$\,kG and $-0.55\pm 0.90$\,kG in this star (i.e. we did not detect a magnetic field). In the present paper, we did not obtain any new field measurements of LMC\,164-2, but we did measure new EWs of the H$\beta$ and H$\alpha$ profiles. 

Using the most recent epochs of OGLE photometry, we derived a new ephemeris for LMC\,164-2, with $T_0=2457506.140$ and $P=7.9606 \pm 0.0009 {\rm d}$. This ephemeris is sufficiently precise to allow us to phase our new EWs with those of \citet{Bagetal17a}. The resultant phase curves, illustrated in Fig.~\ref{Fig_EWs}, appear coherent, but the measurements are too sparse to allow us to evaluate any relationship between the EWs and the stellar brightness in detail. 

From the field measurements obtained in Paper~I, and assuming that the observations were obtained at unknown rotation phases, we estimate that the upper limit of the dipole field is 3.8\,kG. Assuming that the rotation phase is known with respect to the magnetic maximum, the constraint on the dipole field is weaker: There is a 95\,\% probability that the dipolar field strength is between 0 and 6.7\,kG. In any case, our data do not provide evidence that the star is magnetic. We note that the assumption that we know the absolute rotational phase leads to higher \Bupp\ values than in the relative case. One could expect that the former assumption should offer tighter constraints on the \Bupp\ values, rather than looser constraints. In this case, this expectation is not met. For LMC\,164-2, we obtain two measurements that are approximately separated by 0.5 rotational cycles. If we set no constraints on the absolute value of the rotation phase, then we must allow for the possibility that the measurement of $\bz \sim -0.5$\,kG was obtained close to magnetic maximum and the measurement of $\sim +0.2$\,kG was obtained close to the magnetic minimum, a situation that necessarily requires a low \Bdip\ value. If we introduce the constraint that $\bz \sim +0.2$\,kG was obtained close to magnetic maximum and $\bz \sim -0.5$\,kG was obtained close to the magnetic minimum, as we effectively did in the absolute case, then the close-to-zero \Bdip\ values have a lower statistical weight than in the relative case. We should note, however, that the parameter \Bupp\ alone does not suffice to characterise the PDF, and that a comparison between different  \Bupp\  values does not fully describe the difference between PDFs.

The ADM modelling performed by \cite{Munetal20} indicates that the star's surface magnetic dipole should be $B_{\rm dip} = 12.3_{-0.5}^{+1.8}$\,kG, which is higher by a factor of 2 as compared to the upper limit derived here.  The reason for this discrepancy is that  in order to estimate the dipolar field strength, \citet{Munetal20} also used photometric data, a constraint that is not taken into account here. Clearly, more spectropolarimetric data are needed to set more meaningful constraints on the magnetic geometry of LMC\,164-2.

\subsubsection{LMCe\,136-1}
\begin{figure}
\includegraphics*[width=9.4cm,trim={0.8cm 6.0cm 0.3cm 2.5cm},clip]{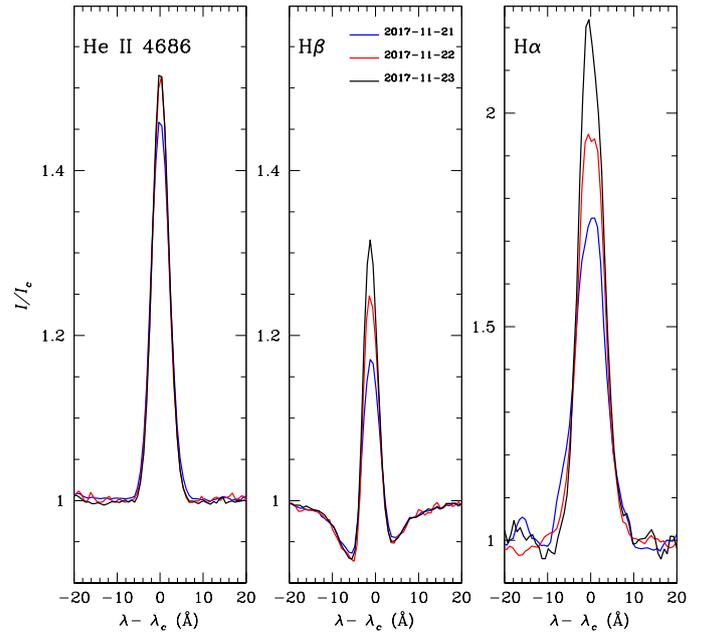}
\caption{\label{Fig_LMCe} Stokes $I$ profiles (normalised to the continuum) for the
spectral lines \HeII\ (left panel), H$\beta$ (mid panel), and H$\alpha$
(right panel) of LMCe\,136-1 observed in three consecutive nights.}
\end{figure}
This new Of?p star ($\alpha$\,$=$\,05:32:22.85, $\delta$\,$=$\,$-$67:10:18.5 in J2000) was found by \citet{Neuetal18}. It was observed in the OGLE~IV survey \citep{Udaetal15}, and photometry was derived with the differential image analysis (DIA) method. We analysed this dataset using the procedures of \citet{Nazetal15} and found a clear period of  $18.706 \pm 0.016$\,d, with $T_0 = 2457503.827$. LMCe\,136-1 is thus an analogue of SMC\,159-1 and LMC\,164-2. The spectropolarimetric data were all obtained at roughly the same phases between maximum and minimum brightness. The EWs measured from these spectra are variable (see Fig.~\ref{Fig_LMCe}), as expected. 

The photometry and EWs as a function of rotation phase are illustrated in Fig.~\ref{Fig_EWs}. As with LMC\,164-2, the EW measurements are too sparse to establish any relationship between the photometric and spectroscopic variations.

The longitudinal field measurements all correspond to null detections ($\bz = 450\pm 235$\,G and $-85\pm 275$\,G).  Taken all together, they do not prove the existence of a magnetic field and set an upper limit on the dipolar field of $\Bupp=4.5$\,kG. This is consistent with the surface magnetic dipole strength inferred by \cite{Munetal20}, $B_{\rm dip} = 5.6_{-1.9}^{+4.1}$~kG.

\subsubsection{AzV\,220}
AzV\,220 was identified as an Of?p star by \citet{Waletal00} and confirmed by \citet{Masduf01}. \citet{Nazetal15} establish photometric variability, but they were unable to derive a period. \citet{Waletal15} report 11 spectra of AzV\,220, confirming clear long-term variability of H$\beta$ and He~{\sc ii}~$\lambda 4686$. In Paper I, we report two new EWs of H$\beta$ and He~{\sc ii}~$\lambda 4686$.
Our new H$\beta$ and H$\alpha$ EWs demonstrate that AzV\,220 continues to exhibit significant and likely complex long-term variability. No period is yet discernible.

Our single measurement of the longitudinal field of AzV\,220 in Paper~I yields a null result with a $\sim 0.6$\,kG uncertainty. In the present paper, we obtain one new polarimetric measurement of AzV\,220: a null result of $0.2 \pm 0.3$\,kG. Altogether, they set an upper limit to the dipolar field strength of $\Bupp=2.3$\,kG. 

\subsection{Main sequence and evolved OBA stars}\label{Sect_OBA}
Both observing campaigns used main sequence and evolved OBA stars to fill unoccupied slitlets.  In the VMS fields, these include
(i) the O supergiants VFTS\,457, Brey\,77 (= Mk\,42), VFTS\,526, and Mk\,37Wa; 
(ii) the B supergiants VFTS 291 and VFTS 458; 
(iii) the O dwarfs  VFTS\,441, VFTS\,500, VFTS\,506, VFTS\,586, and VFTS\,621;
and (iv) the B dwarf VFTS\,589. 
In the Of?p fields, these include 
(v) the B dwarfs AzV\,55 and AzV\,66;
(vi) the A giant NGC\,346 ELS\,19, 
(vii) the OB dwarfs NGC\,346 ELS\,27, NGC\,346 ESL 68, NGC\,346 ELS 100, NGC\,346 ELS\,103, 2dFS\,5037, and 2dFS\,5038; 
and (viii) the giant NGC\,346 1080.
In Table~\ref{Table_Limits}, we have grouped these stars into the following three categories:
OB supergiants, BA giants, and OB dwarfs. The formal uncertainties of their field measurements range from 60 to 1200\,G and dipole field limits are between 1 and 12\,kG.
In the star VFTS\,457, we obtained a field detection ($\bz = 0.61 \pm 0.18$\,kG) on the night of 19\ February 2015, and the analysis of the full dataset provides moderate evidence that the star is magnetic (with an odds ratio of 11). The stellar spectrum is very close to the top edge of CHIP2, and part of the flux of the top beam falls into the gap between the two chips of the CCD, which impedes a correct spectrum extraction. The reliability of our detection is therefore highly questionable. In fact, the analysis of the null field values also provides moderate evidence that the star is magnetic, thus strongly supporting the hypothesis that our field detection is spurious. Therefore, the results for VFTS\,457 have not been included in Table~\ref{Table_Limits}.

Some of our targets were observed with sufficient precision in order to detect typical magnetic fields observed in main sequence stars in the Galaxy. Recent results aimed at measuring magnetic fields in hot Galactic supergiants \citep[e.g.][]{Neietal17,Maretal18} suggest that the fields in those objects are very weak, and it is therefore unsurprising to obtain non-detections in our survey.

\subsection{Cool supergiants}\label{Sect_SG}
The cool luminous supergiants [W60] D24, RM\,1-546, and SkKM\,179 were observed during our Of?p survey; the red supergiant VFTS\,342=WOH\,S\,452 and the yellow supergiant CPD\,$-69$\,463 were observed during our VMS survey. On the star TYC\,8891-3239-1, which was observed during our run dedicated to Of?p stars, we formally detected a magnetic field during the night of 20/21 November 2017 ($\bz=-55\pm15$\,G). Another spectrum was obtained on 22/23 November 2017, but unfortunately it was saturated; therefore, detection could not be confirmed. We must point out that it is unlikely that an uncertainty of 15\,G takes possible spurious effects due to small instrument flexures into account \citep{Bagetal12}; therefore, the star cannot be considered magnetic based only on one measurement. We also note that a 3\,$\sigma$ detection is obtained from the analysis of the null field, further supporting the idea that the field detection is actually spurious. For all cool supergiants, the formal uncertainties  (of the order of 50 to 125\,G) are smaller than for our main targets, and they lead to low upper limits for the dipolar field (of the order of 1\,kG, but as small as 175\,G for [W60]\,D24). On the other hand, measurements of magnetic fields in cool giants and supergiants in the Milky Way \citep[e.g.][]{2010MNRAS.408.2290G} demonstrate that the surface magnetic fields of those stars have longitudinal components smaller than a few Gauss. 

\subsection{WR stars}\label{Sect_WR}
The WR systems VFTS 457, VFTS 507, VFTS 509, VFTS 682, and R136c were observed in the 30 Dor region. No magnetic fields were formally detected with best formal uncertainties for individual stars ranging from 0.07-1~kG. For three WR stars, our Bayesian analysis leads to upper limits of the dipolar field strength of the order of 1-3\,kG. The VMS R136c was observed with a smaller precision leading to an upper limit of $\Bupp = 12.7$\,kG for this star. As field measurements were carried out on emission lines, these estimates refer to the circumstellar environment rather than to the photospheric field.

\subsection{Magnetic wind confinement of very massive stars}\label{subsect:vms_eta}

\begin{table}
\caption[]{Magnetic wind confinement parameters for Of?p and VMS targets. $\eta_*^{95}$ is the value of $\eta_*$ corresponding to $B_{\rm dip}^{\rm (max)}$ in Table 2, and $R_{\rm A}^{95}$ is the corresponding Alfv\'en radius. $P_{\eta_* < 1}$ is the probability that $\eta_*$ is below one. The parameters used to determine $\eta_*$ and $R_*$ were obtained from \citet{2014A&A...570A..38B} for the VMS and from \citet{Munetal20} for the Of?p stars.}
\label{magnetosphere_table}
\begin{tabular}{l r r r r r r}
\hline\hline
Star & $R_*$ & $\log{\dot{M}}$ & $v_\infty$ & $\eta_*^{95}$ & $P_{\eta_* < 1}$ & $R_{\rm A}^{95}$ \\
& ($R_\odot$) & ($M_\odot / {\rm yr}$) & (${\rm km / s}$) & & (\%) & ($R_*$) \\
\hline
\\
\multicolumn{7}{c}{VMS} \\
Mk37Wa & 31 & $-4.6$ & 2530 & 141 & 58 & 3.7 \\
Mk42   & 28 & $-3.9$ & 2840 & 6 & 85 & 1.9 \\
R136c & 37 & $-3.9$ & 1910 & 171 & 57 & 3.9 \\
VFTS 506 & 20 & $-5.6$ & 3040 & 60 & 73 & 3.1 \\
\hline
\\
\multicolumn{7}{c}{Of?p} \\
AzV220 & 10.1 & $-6.5$ & 2330 & 132 & 69 & 3.7 \\
LMC164-2 & 8.4 & $-6.4$ & 2330 & 206 & 61 & 4.1 \\
SMC159-2 & 6.6 & $-6.8$ & 1900 & 2987 & 44 & 7.7 \\
LMCe136-1 & 7.5 & $-6.5$ & 2170 & 330 & 58 & 4.6 \\
\hline\hline
\end{tabular}
\end{table}

\begin{figure}
\includegraphics[width=0.5\textwidth]{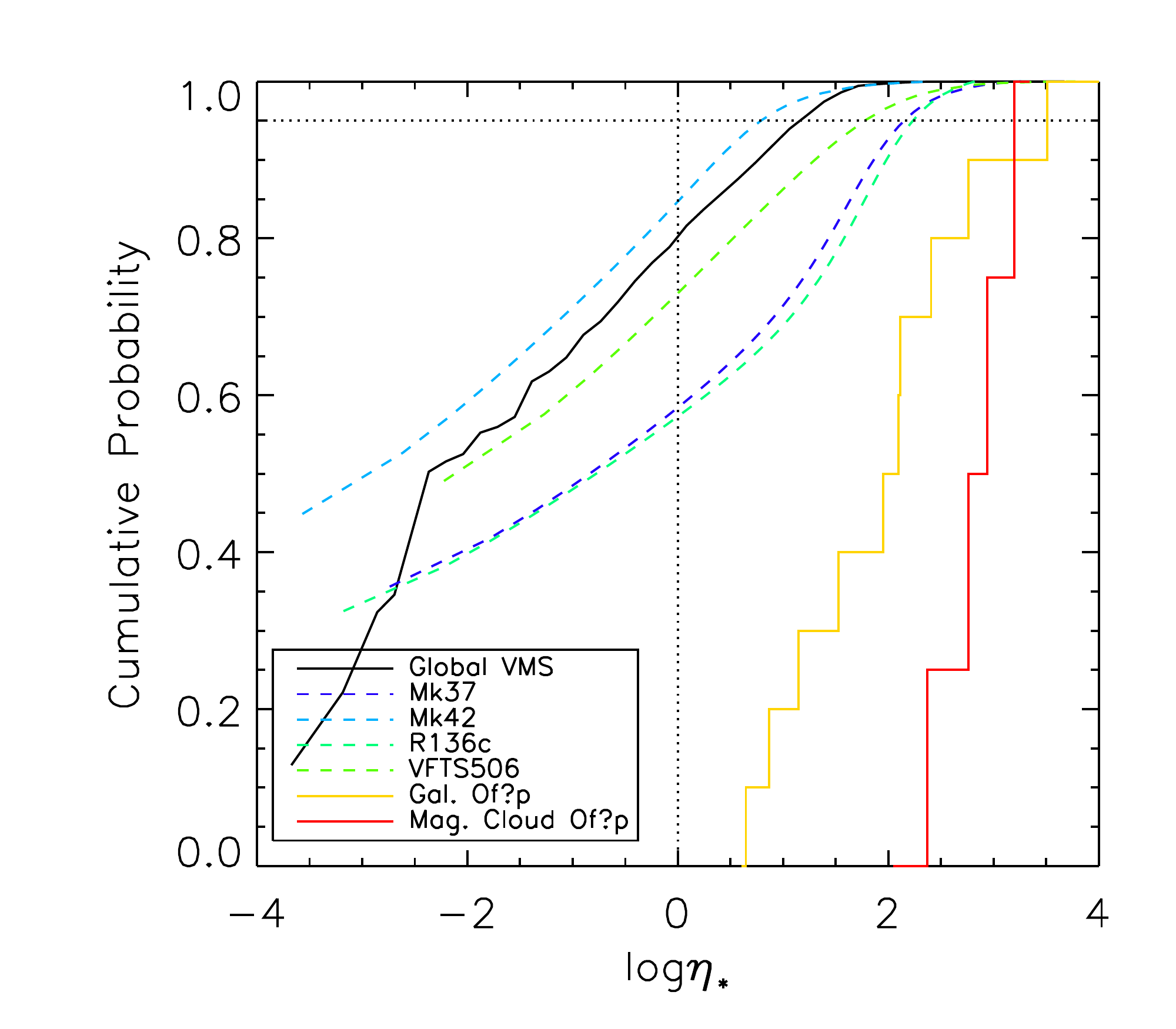}
\caption{\label{vms_eta_cumprob} Cumulative probabilities for $\eta_*$ for individual VMS, Galactic, and extra-galactic Of?p stars (dashed curves), and the global probability for all VMS (solid curve). The vertical and horizontal dotted lines indicate $\eta_* = 1$ and the 95\% confidence level, respectively.}
\end{figure}

Seven of the observed stars are categorised as VMS: Mk37Wa, Mk42, R136c, VFTS 457, VFTS 506, VFTS 621, and VFTS 682. While the upper limits on the magnetic fields of these stars are all in the kG range, given the extremely powerful winds of these stars, it may still be possible that magnetic wind confinement can be ruled out. Since magnetic wind confinement can greatly reduce the net mass-loss rate and hence have a considerable impact on stellar evolution in the upper main sequence \citep[e.g.][]{2017MNRAS.466.1052P} and, therefore, on the mass of the final supernova product, it is of interest to determine whether this phenomenon is relevant to the VMS population.

We determined the magnetic wind confinement parameter $\eta_*$ using Eq.~(7) from \cite{2002ApJ...576..413U}, which also requires the stellar radius, mass-loss rate, and wind terminal velocity. The magnetic wind confinement parameter is simply the ratio of magnetic to kinetic energy density at the stellar surface. If $\eta_*$ is greater than unity, the wind is magnetically confined. These parameters were obtained from \citet{2014A&A...570A..38B}, where radii were determined from the effective temperatures and luminosities given by \citet{2014A&A...570A..38B}. 

We note that \bz~measurements were conducted using emission lines for VFTS 457, 621, and 682; therefore, these measurements probe the circumstellar environment and cannot be used to constrain the photospheric magnetic field (\S~\ref{Sect_WR}; see also Table \ref{Tab_VMS}). Since the calculation of $\eta_*$ requires knowledge of the photospheric magnetic field strength, we excluded these three stars from this analysis. Of the remaining four stars, the spectra of Mk\,37Wa and R136c are blended (Table \ref{Tab_VMS}). However, we have made the assumption that the VMS component dominates the spectrum and that the correction to \bz~for the presence of a second star is negligible. 

The results are given in Table \ref{magnetosphere_table}. The magnetic wind confinement parameter corresponding to $B_{\rm dip}^{\rm (max)}$ in Table~\ref{Table_Limits} (i.e.\ the 95\,\% upper limit for $B_{\rm dip}$) is given by $\eta_*^{95}$. In no case can a magnetically confined wind be ruled out at 95\% confidence. However, the probability that $\eta_*$ is below one is greater than 50\,\% in all cases, and it is almost 85\,\% for Mk\,42. 

The final column of Table \ref{magnetosphere_table} gives the value of the Alfv\'en radius $R_{\rm A}$ corresponding to $\eta_*^{95}$. The Alfv\'en radius is the distance from the star at which the wind opens the magnetic field lines; beyond $R_{\rm A}$, the star no longer has a magnetosphere. We determined $R_{\rm A}$ by using the $\eta_*$ dipole scaling given by \cite{2008MNRAS.385...97U} (their Eq.~(9)). In every case $R_{\rm A}^{95} < 4 R_*$, and in the case of Mk 42, it is 1.9~$R_*$. It is therefore possible that the VMS may have magnetospheres of sizes comparable to those of the Galactic Of?p stars \citep{petit2013}. Following \citeauthor{2008MNRAS.385...97U} (\citeyear{2008MNRAS.385...97U}; see also \citeauthor{2017MNRAS.466.1052P} \citeyear{2017MNRAS.466.1052P}), a reduction in the net mass-loss rate due to magnetic confinement of between about 40\% to 80\% cannot be ruled out. 

In order to examine the wind confinement properties of the population of VMS, we constructed a global PDF by assuming each star contributes probabilistically to various values of $\eta_*$ \citep[as done by][for the case of Galactic O-type stars]{2014MNRAS.444..429D}. The resulting global PDF is shown in Fig.\ \ref{vms_eta_cumprob}. The global PDF indicates that 80\% of the population of VMS should have $\eta_* < 1$, that is,\ the winds of most of the members of this population are not magnetically confined. Also, 95\% of the population should have $\eta_* < 14$. 

For a comparison, we also determined the magnetic wind confinement parameters for the Of?p stars using the upper limits on $B_{\rm dip}$ determined for these stars. Here, we used the fundamental and stellar wind parameters determined for these stars by \cite{Munetal20}. The results are given in Table \ref{magnetosphere_table}. The Of?p stars have, in general, lower probabilities of possessing magnetically unconfined winds, which is reassuring as the winds of these stars must be magnetically unconfined. Correspondingly, the upper limits on $\eta_*$ and $R_{\rm A}$ are significantly higher than those for the VMS. 
A further and more rigorous comparison is provided by examining the values of $\eta_*$ for the Magellanic Cloud Of?p stars inferred by \cite{Munetal20} via ADM modelling as well as the values of $\eta_*$ for Galactic Of?p stars determined directly from magnetic measurements. For the Galactic stars, we used the values published by \cite{petit2013}, with the exception of updated values for $\zeta$ Ori Aa \citep{2015A&A...582A.110B} and HD 108 \citep{2017MNRAS.468.3985S}, and the value determined by \cite{2015A&A...581A..81C} for the magnetic O-type star HD\,54879. The cumulative distributions of the $\log{\eta_*}$ values of Galactic and extra-galactic magnetic O-type stars are shown in Fig.\ \ref{vms_eta_cumprob}. Both are clearly distinct from the VMS sample. In particular, all have $\eta_*$ unambiguously greater than unity. 

While the upper limits on $\eta_*$ for the VMS cannot rule out magnetospheres that are comparable in extent to those of Galactic Of?p stars, it should be emphasised that VMS have not been reported to show the observational signatures of magnetic wind confinement and, in particular, rotational modulation of emission lines. There is, therefore, no evidence that magnetic fields play an important role in the circumstellar environments of these stars. We conclude that it is unlikely that these stars have magnetospheres.

\section{Conclusions}\label{Sect_Conclusions}
We have obtained new magnetic field measurements of over 40 extra-galactic stars. Our main targets were Of?p stars and VMS, but we have also collected several measurements of stars in the vicinity of our main targets by taking advantage of the multi-object capabilities of the FORS2 instrument, which we employed for our observations. By adding up the data presented in this paper with those previously obtained by \citet{Bagetal17a}, we have collected 124 magnetic field measurements of 55 different extra-galactic stars in total. Our final list of observed targets includes five Of?p stars, seven VMS (five O supergiants and two WR stars), and various other giant, supergiant, and WR stars. None of our measurements may be considered as a definite detection, but the results of Bayesian statistical analysis strongly suggest that the Of?p star SMC\,159-2 is magnetic, with a dipolar field strength probably of the order of a 2 to 4\,kG. Our data show no evidence that any of the remaining known Of?p stars of the Magellanic Clouds are magnetic and set upper limits for their dipolar field strength of the order of 3-4\,kG.

We estimated that for the VMS, the typical upper limit on the dipolar field strength is of the order of a few thousand Gauss. Our measurements are therefore unable to probe the regime (a few hundred Gauss) predicted by the magnetic spot model proposed by \cite{CanBra11}. Our results do, however, suggest that not a large fraction of VMS have strong magnetic fields. In fact, the non-detection status is compatible with a magnetic incidence similar to that of Galactic OBA stars, or even lower values. This could be interpreted in three possible ways: either (i) VMS do not form via stellar mergers but, more likely, by classical disc fragmentation, or alternatively (ii) stellar mergers do not always result in the formation of strong magnetic fields, or (iii) the magnetic field decays more quickly in VMS than in canonical massive stars. New measurements of magnetic fields in VMS are needed to put tighter constraints on the possible dipolar strength of VMS, and constrain the formation of these important objects in the Universe.

In the evolved hot stars (luminosity class I and II), because of magnetic flux conservation, we expect very weak fields at the surface, much weaker than for MS stars \citep{Neietal17,Kesetal19}. These fields are in fact very difficult to detect even on Galactic bright stars. Therefore, the fact that we did not detect fields in evolved stars is not surprising. 

\begin{acknowledgements}
  Based on observations made with ESO Telescopes at the La
  Silla Paranal Observatory under programme IDs 094.D-0533 and
  0100.D-0670. We thanks the referee Dr.\ McSwain for a very careful
  review and very useful comments.
  Y.N.\ acknowledges support from the Fonds National de la Recherche
  Scientifique (Belgium) and the University of Li\`{e}ge, as well
  as general support from the PRODEX XMM contract.
  G.A.W.\ acknowledges Discovery Grant support from the Natural Science and
  Engineering Research Council (NSERC) of Canada. 
  ADU gratefully acknowledges the support of NSERC. 
  M.K.Sz.\ acknowledges support from the National Science Centre, Poland,
  grant MAESTRO 2014/14/A/ST9/00121. M.\ E.\ S.\ acknowledges the Annie Jump
  Cannon Fellowship, supported by the University of Delaware and endowed by
  the Mount Cuba Astronomical Observatory.
\end{acknowledgements}
\bibliographystyle{aa}
\bibliography{sbabib.bib}

\begin{appendix}
\section{Observing log and magnetic field measurements}
\begin{table*}[h]
  \caption{\label{Tab_OP} Log of the FORS2 observations obtained with grism
    1200B in spectropolarimetric mode during the run dedicated to Of?p stars (listed in boldface fonts). The first two entries refer to known magnetic Ap stars.}
\begin{tiny}
\begin{tabular}{cccrlrrrrr@{$\pm$}lr@{$\pm$}l}
\hline \hline
DATE  &                    
UT    &                    
\multicolumn{1}{c}{RA}&    
\multicolumn{1}{c}{DEC}&   
\multicolumn{1}{c}{STAR}&  
\multicolumn{1}{c}{$V$} &  
Sp. &                      
Exp &                      
{\it S/N} &                
\multicolumn{2}{c}{\bz} &  
\multicolumn{2}{c}{\nz} \\ 
yyyy-dd-mm & hh:mm & \multicolumn{2}{c}{J2000} & & & & & \AA$^{-1}$ & \multicolumn{2}{c}{G} & \multicolumn{2}{c}{G}\\
\hline
2017-11-21&00:00& 19:53:18.7 &$-$03:06:52&HD\,188041              &$   5.6$&F0 Vp     &   80 &4090 &$  772$&   18&$    4$&   9 \\        
2017-11-21&23:54&            &           &                        &        &         &   80 &2765 &$  779$&   21&$  -21$&  13 \\ [2mm]   
2017-11-21&08:50& 10:55:01.0 &$-$42:15:04&HD\,94660               &$   6.1$&Ap       &   60 &3010 &$-2345$&  25 &$  -15$&  10 \\ [5mm]   
2017-11-22&01:28& 00:49:47.6 &$-$73:17:53& AzV 55                 &$  13.4$&B5       & 6800 &1480 &$ -130$&  60 &$    5$&  60 \\         
2017-11-24&02:25&            &           &                        &        &         & 8880 &1610 &$  -75$&  60 &$ -155$&  60 \\ [2mm]   
2017-11-21&00:32& 00:49:58.7 &$-$73:19:28&{\bf SMC\,159-2}        &$  15.1$&O8f?p    & 7200 & 580 &$  625$& 405& $ -275$& 425 \\         
2017-11-21&04:00&            &           &                        &        &         & 6400 & 540 &$  580$& 505& $  390$& 485 \\         
2017-11-22&01:28&            &           &                        &        &         & 6800 & 605 &$ 1050$& 430& $  865$& 425 \\         
2017-11-24&02:25&            &           &                        &        &         & 8880 & 590 &$  615$& 420& $ 1215$& 440 \\ [2mm]   
2017-11-22&01:28& 00:49:59.7 &$-$73:18:42&2MASS J00495968-7318420 &$  15.2$&HPMS     & 6800 & 675 &$ -185$& 145&$  -35$& 145 \\          
2017-11-24&02:25&            &           &                        &        &         & 8880 & 740 &$ -325$& 170 &$ -95$& 160 \\ [2mm]    
2017-11-22&01:28& 00:50:04.8 &$-$73:21:03&2dFS 5037 &$  15.0$&B0.5 (V)  & 6800 & 635 &$  330$& 260 &$ -540$& 265 \\[2mm]         %
2017-11-22&01:28& 00:50:06.3 &$-$73:16:32& AzV 66                 &$  13.5$&B0.5 V    & 6800 &1380 &$  -75$& 100 &$   55$& 100 \\        
2017-11-24&02:25&            &           &                        &        &         & 8880 &1495 &$ -215$&  95 &$ -285$& 100 \\ [2mm]   
2017-11-22&01:28& 00:50:17.5 &$-$73:17:18&2dFS 5038 &$  15.1$&B0.5 (V)     & 6800 & 715 &$  220$& 190 &$  -55$& 195 \\ [2mm]   %
2017-11-24&04:51& 00:58:53.3 &$-$72:08:35&SkKM 179                &$  12.9$&         G8Iab-Ib     & 8800 & 850 &$ -130$&  50 &$  -15$&  50 \\ [2mm]
2017-11-24&04:51& 00:59:00.9 &$-$72:07:18&NGC 346 ELS 27          &$  15.0$&B0.5 V    & 8800 & 550 &$-1005$& 445 &$ -130$& 460 \\ [2mm]   
2017-11-24&04:51& 00:59:04.2 &$-$72:04:49&NGC 346 ELS 68          &$  15.9$&B0 V(Be-Fe)&8800 & 480 &$ -675$& 830 &$  200$& 920 \\ [2mm]   
2017-11-24&04:51& 00:59:05.6 &$-$72:08:02&NGC 346 ELS 19          &$  14.9$&A0 II     & 8800 & 525 &$ -615$& 315 &$  685$& 385 \\ [2mm]   
2017-11-24&04:51& 00:59:10.0 &$-$72:05:49&{\bf AzV\,220}          &$  14.5$&O6.5f?p  & 8800 & 845 &$ -225$& 305 &$ -155$& 290 \\ [2mm]   
2017-11-24&04:51& 00:59:18.3 &$-$72:04:21& NGC 346 1080    &$  16.1$&B0.5 III  & 8800 & 535 &$  475$& 395 &$  575$& 425 \\ [2mm]  
2017-11-24&04:51& 00:59:20.8 &$-$72:02:59&NGC 346 ELS 103         &$  16.2$&B0.5 V    & 8800 & 430 &$ -320$& 480 &$  630$& 490 \\ [2mm]   
2017-11-24&04:51& 00:59:20.8 &$-$72:03:38&NGC 346 ELS 100         &$  16.1$&B1.5 V    & 8800 & 430 &$  -50$& 595 &$  190$& 575 \\ [2mm]   
2017-11-21&06:40& 05:31:09.9 &$-$67:11:19&[W60] D24               &$  13.2$& M1 I    & 9600 &1065 &$    5$&  30 &$  -30$&  30 \\         
2017-11-23&07:39&            &           &                        &        &         & 7200 &1050 &$    5$&  30 &$   15$&  30 \\ [2mm]   
2017-11-21&06:40& 05:31:54.5 &$-$67:08:22&RM 1-546                &$  13.4$& M: E    & 9600 & 695 &$   10$&  50 &$   70$&  50 \\         
2017-11-23&07:39&            &           &                        &        &         & 7200 & 655 &$  100$&  45 &$  -85$&  50 \\ [2mm]   
2017-11-21&06:40& 05:31:54.8 &$-$67:11:21&2MASS 05315473-6711194  &$  14.7$&         & 9600 & 845 &$ -395$& 200 &$  225$& 205 \\         
2017-11-23&07:39&            &           &                        &        &         & 7200 & 820 &$ -220$& 200 &$  350$& 205 \\ [2mm]   
2017-11-21&06:40& 05:32:22.9 &$-$67:10:19&{\bf LMCe\,136-1}       &$  14.6$&O6.5f?p  & 9600 & 995 &$  450$& 235 &$  -10$& 235 \\         
2017-11-23&07:39&            &           &                        &$      $&         & 7200 & 890 &$  -85$& 275 &$ -230$& 280 \\ [2mm]   
2017-11-21&06:40& 05:32:32.9 &$-$67:10:42&TYC 8891-3239-1         &$  11.3$&         & 9600 &3085 &$  -55$&  15 &$  -45$&  15 \\ [2mm]   
2017-11-21&06:40& 05:32:39.2 &$-$67:09:49&2MASS 05323916-6709487  &$  14.4$&         & 9600 &1095 &$  375$& 110 &$  -25$& 110 \\         
2017-11-23&07:39&            &           &                        &        &         & 7200 &1060 &$  130$& 105 &$  135$& 105 \\ [2mm]   
2017-11-21&06:40& 05:32:47.4 &$-$67:10:05&SK-67 180               &$  12.6$& B3: I   & 9600 &2280 &$   -5$&  50 &$   30$&  50 \\         
2017-11-23&07:39&            &           &                        &        &         & 7200 &2260 &$ -125$&  55 &$   85$&  50 \\ [2mm]   
2017-11-21&06:40& 05:32:55.2 &$-$67:10:20&2MASS 05325508-6710200  &$  15.7$& ??      & 9600 & 580 &$ -180$& 325 &$ -535$& 325 \\         
2017-11-23&07:39&            &           &                        &        &         & 7200 & 575 &$  320$& 180 &$   20$& 280 \\ [2mm]   

\hline
  \end{tabular}
\end{tiny}

\noindent
\tablefoot{\small
Columns~1 and 2 give the civilian date and UT time of the midpoint of the observation; columns~3--5 give the J2000
RA and DEC and the target name as identified through SIMBAD or other catalogues;
col.~6 gives the $V$ magnitude;
col.~7 is the star's spectral type;
cols.~8 and 9 give the total exposure time and the {\it S/N} per \AA; and
cols.~10 and 11 give our field determination from the reduced Stokes $V$ profiles, \bz, and from the null profiles, \nz, respectively.
}
\end{table*}
\begin{table*}
  \caption{\label{Tab_VMS}  Log of the FORS2 observations from the run on very massive stars (VMS)
  and field measurements from absorption lines.
    Columns as in Table~\ref{Tab_OP}, except that exposure time is not reported 
    as it was identical (3296\,s) for all observations. Col.~5 gives two designations
    of the same target. Col.~11 refers to the comments at the end of the Table.
    VMS names are typed with boldface fonts.}
    \tabcolsep=0.14cm
\begin{tiny}
\begin{center}
\begin{tabular}{cccrlrrrr@{$\pm$}lr@{$\pm$}lr}
\hline \hline
DATE  &                    
UT    &                    
\multicolumn{1}{c}{RA}&    
\multicolumn{1}{c}{DEC}&   
\multicolumn{1}{c}{STAR}&  
\multicolumn{1}{c}{$V$} &  
Sp. type&                  
{\it S/N} &                
\multicolumn{2}{c}{\bz} &  
\multicolumn{2}{c}{\nz} &  
Note                      \\
yyyy-dd-mm & hh:mm & \multicolumn{2}{c}{J2000} & & & & \AA$^{-1}$ & \multicolumn{2}{c}{G} & \multicolumn{2}{c}{G}\\
\hline
2015-03-05&01:37&05:38:17.7&$-69$:03:39&VFTS 291               &  14.9  &B5\,II-Ib        & 395 &$  -75$&  770&$ 1035$&  765& \\        
2015-03-05&03:01&          &           &[HBC93] 275            &        &                 & 355 &$  910$&  980&$ -815$& 1005& \\ [2mm]  
2015-02-19&03:38&05:38:26.7&$-69$:08:53&VFTS 341               &$ 14.0 $&K5Ia-Iab         & 560 &$  105$&  145&$  200$&  150& \\        
2015-02-25&03:14&          &           &WOH S 452              &        &                 & 570 &$ -185$&  125&$  -20$&  130& \\        
2015-03-05&01:37&          &           &                       &        &                 & 540 &$  390$&  145&$   30$&  135& \\        
2015-03-05&03:01&          &           &                       &        &                 & 490 &$   60$&  170&$  210$&  175& \\        
2015-04-01&00:51&          &           &                       &        &                 & 545 &$  135$&  130&$  -75$&  145& \\ [2mm]  
2015-02-19&03:38&05:38:37.8&$-69$:03:29&VFTS 441               &$ 15.1 $&O9.5V            & 340 &$-1660$& 1025&$ -840$&  960& \\        
2015-02-25&03:14&          &           &[P93] 613              &        &                 & 360 &$  405$& 1110&$ -900$& 1100& \\        
2015-04-01&00:51&          &           &                       &        &                 & 375 &$ -585$& 1140&$ -555$& 1110& \\ [2mm]  
2015-02-19&03:38&05:38:38.8&$-69$:06:50&{\bf VFTS 457}         &$ 13.7 $&O3.5\,If*/WN7    & 645 &$  605$&  175&$   40$&  100&1 \\       
2015-02-25&03:14&          &           & Mk 51                 &        &                 & 640 &$   20$&  140&$  445$&  140&  \\       
2015-03-05&01:37&          &           &                       &        &                 & 710 &$ -600$&  310&$  590$&  285&  \\       
2015-03-05&03:01&          &           &                       &        &                 & 620 &$ -795$&  340&$  780$&  280&  \\       
2015-04-01&00:51&          &           &                       &        &                 & 630 &$  190$&  140&$  -65$&  135&  \\ [2mm]
2015-03-05&01:37&05:38:38.9&$-$69:08:15&  VFTS 458             &  12.6  &BN6Iap           & 995 &$  165$&  285&$ -180$&  260& \\        
2015-03-05&03:01&          &           &  HTR 13               &        &                 & 895 &$ -350$&  360&$  645$&  360& \\ [2mm]  
2015-03-05&01:37&05:38:41.2&$-69$:02:58&VFTS 500            &14.2&O6.5IV((fc))+O6.5V((fc))& 590 &$  790$&   650&$ 1390$&  575& \\       
2015-03-05&03:01&          &           &LMC 171520             &        &                 & 515 &$  970$&   655&$-1880$&  640& \\ [2mm] 
2015-02-19&03:38&05:38:41.5&$-69$:05:20&{\bf VFTS 506}         &$ 13.1  $&ON2V((n))((f*)) & 750 &$  -50$&  720&$ 825 $&  690 & \\       
2015-02-25&03:14&          &           &NGC 2070 Mk 25         &         &                & 790 &$  340$&  705&$ -915$&  695 & \\       
2015-04-01&00:51&          &           &                       &        &                 & 795 &$ -450$&  670&$  225$&  620 & \\ [2mm] 
2015-03-05&01:37&05:38:41.6&$-69$:05:13& VFTS 507              &  12.5  &WC4+WN6+O        &2220 &$  125$&  165&$ -260$&  190 &1,2\\     
2015-03-05&03:01&          &           & R140a                 &        &                 &2245 &$   30$&  160&$  -85$&  155 &   \\ [2mm]
2015-03-05&01:37&05:38:41.6&$-69$:05:14&VFTS 509               &$ 12.8 $&WN6+O            &1120 &$  -60$&  210 &$  70$& 180 &1,2\\      
2015-03-05&03:01&          &           &R140b                  &        &                 & 880 &$   20$&  300 &$-430$& 270 &   \\ [2mm]
2015-02-19&03:38&05:38:41.9&$-69$:06:13&{\bf Mk\,37Wa}         &$ 14.9 $&O4If+            &  845&$  285$&  735&$ -905$&  685&2\\        
2015-02-25&03:14&          &           &VFTS 1021              &        &                 &  860&$-1385$&  640&$ 1320$&  580& \\        
2015-04-01&00:51&          &           &                       &        &                 &  910&$ -730$&  545&$  760$&  505& \\ [2mm]  
2015-03-05&01:37&05:38:42.1&$-69$:05:55&{\bf Mk\,42}           &  11.0  & O2If*           &1175 &$ -645$&  430&$  195$&  360& \\        
2015-03-05&03:01&          &           &  BAT99 105            &        &                 &1075 &$ -220$&  465&$ -135$&  395& \\ [2mm]  
2015-02-19&03:38&05:38:42.2&$-69$:08:32&  VFTS 526             &$ 15.2 $&O8.5I((n))fp     & 350 &$-1195$&  425&$ -530$&  425& \\        
2015-02-25&03:14&          &           & [P93] 925             &        &                 & 350 &$-1055$&  470&$-1155$&  480& \\        
2015-04-01&00:51&          &           &                       &        &                 & 340 &$ -330$&  505&$  -85$&  425& \\ [2mm]  
2015-02-19&03:38&05:38:42.7&$-69$:06:04&{\bf R136c}            &$      $&WN5h             & 1440&$ 1065$& 1290&$-1030$& 1120&2\\        
2015-02-25&03:14&          &           &VFTS 1025              &        &                 & 1200&$ 1230$& 1145&$ -100$& 1110& \\        
2015-04-01&00:51&          &           &                       &        &                 & 1275&$ 2745$& 1300&$-1480$& 1300& \\ [2mm]  
2015-02-19&03:38&05:38:45.4&$-69$:02:51&VFTS 586               &$ 14.5 $&O4V((n))((fc))z  & 370 &$ 1485$& 1085&$ 1645$& 1020& \\        
2015-02-25&03:14&          &           &                       &       &                  & 365 &$ -820$&  690&$ -965$&  775& \\        
2015-04-01&00:51&          &           &                       &        &                 & 375 &$ 1205$&  845&$ 1175$&  870& \\ [2mm]  
2015-03-05&01:37&05:38:45.6&$-68$:07:35&  VFTS 589             &  15.8  &B0.5V(SB2?)      & 270 &$ 1585$& 1205&$  255$& 1110& \\        
2015-03-05&03:01&          &           &  [P93] 1247           &        &                 & 210 &$  745$& 1785&$ 2185$& 1775& \\ [2mm]  
2015-02-19&03:38&05:38:48.1&$-69$:04:42&{\bf VFTS 621}         &$ 15.4  $&O2V((f*))z      & 415 &$   70$&  285&$  -35$& 195 &1\\        
2015-02-25&03:14&          &           &Walborn 2              &        &                 & 465 &$ -405$&  330&$ -420$& 365 & \\        
2015-04-01&00:51&          &           &                       &        &                 & 525 &$ -100$&  700&$  100$& 515 & \\ [2mm]  
2015-02-19&03:38&05:38:51.6&$-69$:08:07& CPD-69 463            &$ 12.0 $&F7Ia             &1535 &$ -170$&  185&$  -55$&  170& \\        
2015-02-25&03:14&          &           & R143                  &        &                 &1570 &$  125$&  155&$  180$&  160& \\        
2015-04-01&00:51&          &           &                       &        &                 &1540 &$ -265$&  180&$  -10$&  185& \\ [2mm]  
2015-03-05&01:37&05:38:55.5&$-69$:04:27&{\bf VFTS 682}         &  16.1  &WN5h             & 230 &$  -30$&  650&$  145$&  665& 1\\       
2015-03-05&03:01&          &           & LMCe 1415             &        &                 & 200 &$  185$&  755&$ -160$&  770& 1\\ [2mm] 
\hline
  \end{tabular}
\end{center}
\end{tiny}
\tablefoot{
\noindent
{\bf 1:} Magnetic field was estimated from emission lines. 
{\bf 2:} Spectrum blended.
}
\end{table*}

\end{appendix}
\end{document}